\documentclass[usenatbib,useAMS]{mn2e}
\usepackage{times}
\usepackage{epsfig}

\renewcommand{\vec}[1]{\bmath{#1}}

\newcommand{\aries}{\gamma}

\begin{document}

\title[Planetary microlensing from orbital source motion]{Planetary microlensing signals from the orbital motion of the source star around the common barycentre}
\author[S. Rahvar and M. Dominik]
{S. Rahvar$^{1}$ and M. Dominik$^{2}$\thanks{Royal Society University Research Fellow}  \\
$^1$Department of Physics, Sharif University of Technology,
 P.O.~Box 11365--9161, Tehran, Iran\\
$^{2}$SUPA, University of St Andrews, School of Physics \&
Astronomy, North Haugh, St Andrews, KY16 9SS, United Kingdom}

 \maketitle





\begin{abstract}
With several detections, the technique of gravitational microlensing 
has proven useful for studying planets that orbit stars at Galactic
distances, and it can even be applied to detect planets in neighbouring
galaxies. So far, planet detections by microlensing have been considered
to result from a change in the bending of light and the resulting
magnification caused by a planet around the foreground lens star.
However, in complete analogy to the annual parallax effect caused by the
revolution of the Earth around the Sun, the motion of the source 
star around the common barycentre with an orbiting planet can also lead
to observable deviations in microlensing light curves that can 
provide evidence for the unseen companion. We discuss this effect
in some detail and study the prospects of microlensing observations
for revealing planets through this alternative detection channel.
Given that small distances between lens and source star are favoured,
and that the effect becomes nearly independent of the source distance,
planets would remain detectable even if their host star is located outside the Milky Way with a sufficiently good photometry (exceeding present-day 
technology) being possible.
From synthetic light curves arising from a Monte-Carlo simulation, we
find that the chances for such detections are not overwhelming and appear
practically limited to the most massive planets (at least with current
observational set-ups), but they are large enough for leaving the
possibility that one or the other signal has already been
observed. However, it may remain undetermined whether
the planet actually orbits the source star or rather the lens star, which
leaves us with an ambiguity not only with respect to its location, but
also to its properties.


\end{abstract}

\section{Introduction}
Gravitational microlensing, i.e.\ the transient brightening of an
observed star due to the bending of light caused by the gravitational
field of an intervening foreground 'lens' star, was considered by Einstein
as early as 1912, as pointed out by \citet{RSS}, but he concluded
that "there is no great chance of observing this phenomenon"
\citep{ein36}. Only several decades of advance in technology enabled
the first reported discovery of a microlensing event \citep{alc93},
following the suggestion by \citet{pac86} to use the technique as a tool for detecting compact matter in the Galactic halo.


However, microlensing provides a valuable tool for a variety of 
other astrophysical applications, and the most spectacular one nowadays
is the detection of extra-solar planets. It was already pointed out by
\citet{Lie64} that the primary effect of planets as gravitational lenses
would be to produce a slight perturbation of the lens action of their
respective host star. The distortion of the magnification pattern of
the foreground lens star by an orbiting planet, and the additional short blip or dip to the otherwise symmetric microlensing light curve was then 
first studied by \citet{mao91}. A super-Jupiter
was the first planet detected by this technique \citep{bon04}, but
its sensitivity even reaches below the mass of Earth, even for
ground-based observations \citep{ben96,dom07}. In fact,
the detectability of planets below 10 $M_\oplus$ has been 
impressively demonstrated with the first discovery of a cool rocky/icy exoplanet \citep{bea06}. Microlensing is already singled out
amongst all ground-based current techniques aiming at the detection of extra-solar
planets by the respective host stars being at Galactic distances, rather
than in the solar neighbourhood, and
belonging to either of two stellar populations. Significantly beyond this,
even planets orbiting stars in neighbouring galaxies, such as M31, 
could be detected \citep{cov00,chu06}, whereas no other technique so
far has been suggested that could achieve such a goal within foreseeable
time.

By creating a link between received light, the gravitational field of intervening objects, and relative transverse motions
between source, lens, and observer, the effect of gravitational microlensing shows a substantial versatility. It is therefore not that surprising that it provides other channels for detecting planets
orbiting stars other than the Sun. As suggested by \citet{GG00}, 
the light of close-in gas-giant planets would be detectable with large
telescopes if such observations are scheduled while the planet follows
its host star in exiting a caustic produced by a binary lens system,
given that the light received from the planet would be far more strongly magnified than that received from its host star.

Here, we discuss a further channel for revealing the existence of
extra-solar planets from the study of microlensing light curves.
Rather than considering a planet around the lens star, we study the
effects of the orbital motion of the source star and a planetary
companion around the common barycentre. While the planet is not seen
itself, the small motion of its observed host star periodically alters
the line-of-sight and thereby the relative lens-source position, which
results in a change of the observed source magnification. In fact,
an analogous effect is caused by the revolution of the Earth around the
Sun, where the line-of-sight is altered due to the motion of the observer
rather than the observed object, so that it constitutes a form of
parallax effect. While the annual parallax in a microlensing event
was first observed by \citet{alc95}, the orbital motion of stellar
source binaries has been studied extensively by various authors \citep{gri92,han97,pac97}. Some authors refer to the latter as
'xallarap' effect, but in fact, this nomenclature involves a double inversion, because parallax is known as the apparent change in position of an observed object that is just the reflection of a change in position by the observer.
Not surprisingly, both effects are not easy to distinguish, and 
a systematic analysis of 22 microlensing parallax candidate events
\citep{poi05} explicitly finds 23 per cent of them being strongly affected by 'xallarap'. Future space missions capable of accurate astrometry
such as SIM and GAIA will however allow measure the parallax along with
the relative proper motion between lens and source star, allowing to
obtain more accurate information about the Galaxy 
\citep*[e.g.][]{pac98,rahvar2005}.

We start our discussion in Sect.~\ref{sec:orbits} with an introduction to the
 basics of gravitational
microlensing and Keplerian motion, followed by a review of the
annual Earth-Sun parallax based on the common formalism for
orbital motion in observer, lens, and source developed by
\citep{Do:rotate}. The discussion of annual parallax is subsequently
transferred to the case of stellar reflex motion due to an orbiting
planet, where the relevant parameters are identified.
In Sect.~\ref{sec:detect}, we turn our attention to the prospects 
for planet detection through our proposed channel, where we first
discuss the strength of the observable effect and identify the favourable
scenarios. Subsequently, we discuss the input parameters and the results
of a Monte-Carlo simulation of synthetic light curves, thereby quantifying
the planet detectability as a function of the planet mass and orbital parameters. Moreover, we study how to extract properties of the planet
and its orbit from the observed light curve. A short summary and
final conclusions are presented in Sect.~\ref{sec:conclusions}.

\section{Planetary orbits and microlensing}
\label{sec:orbits}

\subsection{Microlensing events}
According to the theory of General Relativity, a light ray passing
beside a massive object is bent, the arrival time of its photons is
delayed, and a ray bundle is distorted. Allowing for several
possible light rays from an observed source object to the observer,
this phenomenon is commonly known as `gravitational lensing'. If a
foreground star with mass $M$ at distance $D_\rmn{L}$ happens to be
sufficiently aligned with an observed background star at distance
$D_\rmn{S}$, the time delay is negligible, while the angular
separation between the images is of the order of micro-arcseconds,
and therefore undetectable with current telescopes. However,
contrary to extragalactic scenarios of gravitational lensing, stars
in our own or neighbouring galaxies show a substantial proper
motion, so that the image distortion results in an observable transient
brightening as the foreground `lens' star passes along the
line-of-sight to the observed background source star, which is
commonly referred to as a `microlensing event'.

Explicitly, for lens and source being separated on the sky by a
position angle $\vec u\,\theta_\rmn{E}$, where the angular Einstein radius
\begin{equation}
\theta_\rmn{E} = \sqrt{\frac{4GM}{c^2}\,\left(D_\rmn{L}^{-1}
-D_\rmn{S}^{-1}\right)}
\label{eq:thetaE}
\end{equation}
provides the unique characteristic scale of gravitational microlensing,
the combined magnification of the images reads \citep{ein36}
\begin{equation}
A(u) = \frac{u^2 + 2}{u\sqrt{u^2 +4}}\,,
\label{eq:magpoint}
\end{equation}
where $u = |\vec u|$.

Let us consider source and lens star being in uniform proper motion
with $\vec \mu_\rmn{S}$ and $\vec \mu_\rmn{L}$, respectively, so that
the relative proper motion reads $\vec \mu = \vec \mu_\rmn{S}-\vec \mu_\rmn{L} = \mu\,(\cos \psi,\sin \psi)$. With $u_0$ marking the
closest angular approach between lens and source star, realized at
epoch $t_0$, the relative source-lens trajectory takes the form
\begin{equation}
\vec u_\rmn{LS}(t) = u_0\,\left(\begin{array}{c} -\sin\psi \\
\cos\psi\end{array}\right) + \omega_\mathrm{E}\,(t-t_0)\,
\left(\begin{array}{c} \cos\psi \\
\sin\psi\end{array}\right)\,,
\label{eq:sourcemotion}
\end{equation}
where $\omega_\rmn{E} \equiv \mu/\theta_\rmn{E}$, so that
\begin{equation}
u_\rmn{LS}(t) = \sqrt{u_0^2 + \omega_\rmn{E}^2\,(t-t_0)^2}\,.
\end{equation}
In the literature, rather than $\omega_\rmn{E}$,
a time-scale $t_\rmn{E} \equiv \omega_\rmn{E}^{-1}$ is used more frequently.
Given that $A(u)$ is strictly decreasing with $u$, and $u(t)$ assumes
a minimum of $u_0$ at epoch $t_0$, the magnification $A[u(t)]$ leads
to a symmetric light curve, peaking at $t_0$ \citep{pac86}.

The symmetry of the light curve is retained if the finite angular radius
$\theta_\star$ of the source star is taken into account. With a source
size parameter $\rho_\star = \theta_\star/\theta_\rmn{E}$ and the
abbreviations
\begin{equation}
n = \frac{4\,u\,\rho_\star}{(u+\rho_\star)^2}\,, \quad
k = \sqrt{\frac{4\,n}{4+(u-\rho_\star)^2}}\,,
\end{equation}
\citet{WittMao} found a uniformly bright star being magnified by
\begin{eqnarray}
A(u; \rho_\star) &  = & \frac{1}{2\upi}\,\left[\frac{u+\rho_\star}{\rho_\star^2}\,\sqrt{4+(u-\rho_\star)^2}\,E(k)\right.\;-\nonumber \\
& & \; -\;\frac{u-\rho_\star}{\rho_\star^2}\,\frac{8+u^2-\rho_\star^2}
{\sqrt{4+(u-\rho_\star)^2}}\,K(k)\;+\nonumber \\
& & \; +\;\left.\frac{4\,(u-\rho_\star)^2}{\rho_\star^2\,(u+\rho_\star)}\,\frac{1+\rho_\star^2}{\sqrt{4+(u-\rho_\star)^2}}\,\Pi(n;k)\right]
\end{eqnarray}
for $u \neq \rho_\star$, where $K(k)$, $E(k)$, and 
$\Pi(n; k)$ denote the complete elliptic integrals of first,
second, and third kind, respectively, whereas
\begin{equation}
A(u;\rho_\star) = \frac{1}{\upi}\left[\frac{2}{\rho_\star} +
\frac{1+\rho_\star^2}{\rho_\star^2}\,\left(\frac{\upi}{2} +
\arcsin \frac{\rho_\star^2-1}{\rho_\star^2+1}\right)\right]
\end{equation}
for $u = \rho_\star$. The size of the source star significantly affects
the observed light curve for $u \la 2\,\rho_\star$, while
$A(u;\rho_\star)$ reveals Eq.~(\ref{eq:magpoint}) as $\rho_\star \to 0$.

For $F_\rmn{S}$ denoting
the intrinsic flux of the magnified source star and $F_\rmn{B}$ that of the
background, unresolved from the observed target,
the microlensing light curve is given by the received flux
\begin{equation}
F(t) = F_\rmn{S}\,A[u(t);\rho_\star,I] + F_\rmn{B}\,.
\end{equation}
where the finite-source magnification $A[u(t),\rho_\star]$ is completely characterized by $(u_0,t_0,\omega_\rmn{E},\rho_\star,I)$, where $I$ stands
for the brightness profile function of the source star, for which we
assume $I \equiv 1$ (i.e.\ a uniformly bright source) throughout.

\subsection{Keplerian motion}
If one neglects relativistic effects, a planet and its host star
of masses $m_\rmn{p}$ and $m_\star$, respectively,
both are in elliptic orbits around their common centre of mass.
In fact, the motion of their separation vector $\vec r(t)
= \vec r_\rmn{p}(t)- \vec r_\star(t)$ can be
understood as a fixed virtual body of total mass
\begin{equation}
m = m_\rmn{p}+m_\star
\end{equation}
being orbited by another virtual body of the reduced mass
\begin{equation}
\mu = \frac{m_\rmn{p}\,m_\star}{m}\,,
\end{equation}
and with $\vec r_\rmn{cm}(t)$ denoting the motion of the centre of mass,
\begin{eqnarray}
\vec r_\rmn{p}(t) &=& \vec r_\rmn{cm}(t) + (m_\star/m)\,\vec r(t) \nonumber \\
\vec r_\star(t) &=& \vec r_\rmn{cm}(t) - (m_\rmn{p}/m)\,\vec r(t)
\label{eq:barymotion}
\end{eqnarray}

The orbit $\vec r(t) = [x(t), y(t), 0]$ is characterized by its major semi-axis $a$, its orbital period $P$, its eccentricity $\varepsilon$, the orbital plane,
spanned by $x$ (along major axis) and $y$ (along minor axis),
as well as the orbital phase at a reference epoch. Kepler's third law provides a relation between the
orbital period $P$ and the major semi-axis $a$ by means of the
total mass $m$, where
\begin{equation}
P = 2\,\pi\,\sqrt{\frac{a^3}{Gm}}\,.
\label{eq:kepler3}
\end{equation}
With positive $x$ in the direction of periastron
and positive $y$ from periastron towards the motion of the planet,
one finds
\begin{eqnarray}
x(t) &= & a \left[\cos \xi(t) - \varepsilon\right]\,,\nonumber \\
y(t) &=& a\,\sqrt{1-\varepsilon^2}\,\sin \xi(t)\,,
\end{eqnarray}
where the eccentric anomaly $\xi \in [0,2\,\upi)$ is given by
\begin{equation}
\xi(t) - \varepsilon\,\sin \xi(t)
= \Omega(t-t_\rmn{p})-\lfloor \Omega(t-t_\rmn{p}) \rfloor
\,,
\end{equation}
with $\Omega = (2\,\upi)/P$, and
$t_\mathrm{p}$ being an epoch of periastron.

For small eccentricities, an expansion in the lowest order
of the eccentricity $\varepsilon$ is a fair approximation, which reads
\begin{eqnarray}
x(t) &=& a\,\rho(t)\,\cos \xi(t)\,,\nonumber\\
y(t) &=& a\,\rho(t)\,\sin \xi(t)\,,
\label{eq:smalleps}
\end{eqnarray}
with
\begin{equation}
\rho(t) = 1-\varepsilon\,\cos\left[\Omega(t-t_\rmn{p})\right]
\end{equation}
and
\begin{equation}
\xi(t) = \Omega(t-t_\rmn{p}) + 2\varepsilon\,\sin\left[\Omega(t-t_\rmn{p})\right]\,.
\label{eq:xi}
\end{equation}

\subsection{Earth-Sun parallax}
For the Earth, $m_\rmn{p} \ll m_\star$, and with the Sun at rest,
$\vec r_\rmn{p} \approx \vec r(t)$. Moreover, the orbital major semi-axis
is given by $a = 1~\mbox{AU}$, the orbital frequency is given by
$\Omega = (2\upi)/(1~\mbox{yr})$, and the eccentricity $\varepsilon = 0.017$ is small. The effect of the annual parallax, i.e.\ the revolution of the Earth around the Sun has been discussed in some detail by
\citet{Do:rotate}, and our subsequent discussion is based on the
results derived in that context.

The (parallactic) shift of the position of the observer
 $(\delta r)_\rmn{O}$ perpendicular to the line-of-sight is
equivalent to a virtual displacement of the observed source star $(\delta r)_\rmn{S}$. For the corresponding shift in the dimensionless angular
coordinate $(\delta u)_\rmn{S} = (\delta r)_\rmn{S}/(D_\rmn{S}\,\theta_\rmn{E})$, one finds
\begin{equation}
(\delta u)_\rmn{S} = \frac{D_\rmn{S}-D_\rmn{L}}{D_\rmn{L}\,D_\rmn{S}}\,
\frac{(\delta r)_\rmn{O}}{\theta_\rmn{E}} =
\left(\frac{1}{D_\rmn{L}} - \frac{1}{D_\rmn{S}}\right)\,
\frac{(\delta r)_\rmn{O}}{\theta_\rmn{E}}\,.
\end{equation}
With the relative lens-source parallax
\begin{equation}
\pi_{\rmn LS} = 1~\mbox{AU}\,\left(\frac{1}{D_\rmn{L}} - \frac{1}{D_\rmn{S}}\right)\,,
\end{equation}
it is customary to define a microlensing parallax
$\pi_\rmn{E} = \pi_{\rmn LS}/\theta_\rmn{E}$\,.

A standard coordinate system based on the Earth's orbital plane is given
by the ecliptical coordinates $(\beta,\lambda)$, where positive $\beta$
point towards ecliptical North, while increasing $\lambda$ follow the
apparent motion of the Sun, which is in the same sense as the actual
motion of the Earth. Moreover, $\lambda = 0$ is defined as the position
of the Sun at vernal equinox.
In fact, the ecliptical latitude $\beta$ equals the inclination of the
Earth's orbit with respect to the line-of-sight to a source star at
ecliptical coordinates $(\beta, \lambda)$.

In order to determine the effect of the orbital motion of the Earth,
one needs to find its components perpendicular to the line-of-sight.
While $(x,y)$ span the orbital plane, where the positive x-axis points
towards perihelion, the complementary $z$-coordinate points towards
ecliptical North, thereby forming a right-handed three-dimensional system.
We can also define a coordinate system with $(\hat u_1,\hat u_2)$
in the plane
perpendicular to the line-of-sight and $\hat u_3$ pointing towards the
observer, so that again a right-handed system is formed. Rather than
using $\lambda$ as the longitude, it is more straightforward to choose an angle $\varphi$, where $\varphi = 0$ corresponds to the Earth being at the perihelion. In fact, $\varphi = \lambda + \upi + \varphi_\aries$, where $\varphi_\aries$ is the longitude of the
vernal equinox as measured from the perihelion. While we can define
$\hat u_1 \propto z$, $\hat u_2 \propto y$, and $\hat u_3 \propto -x$ for $\beta = 0$
and $\varphi = 0$, the coordinates for arbitrary $(\beta,\varphi)$ arise
from a rotation around the $\hat u_1$-axis by the angle $\varphi$, and a
subsequent rotation around the $\hat u_2$-axis by the angle $-\beta$,
so that
\begin{equation}
\left(\begin{array}{c}
\hat u_1(t) \\ \hat u_2(t)
\end{array}\right)
= \frac{\pi_\rmn{E}}{1~\mbox{AU}}\,{\cal R}(\beta,\varphi)\,
\left(\begin{array}{c}
x(t) \\ y(t)
\end{array}\right) \,,
\end{equation}
where
\begin{equation}
{\cal R}(\beta,\varphi) = \left(\begin{array}{cc}
-\sin \beta\,\cos\varphi & -\sin \beta\,\sin\varphi \\
-\sin \varphi & \cos\varphi
\end{array}
\right)\,.
\label{eq:rotmatrix}
\end{equation}
Since, in the small-eccentricity limit, Eq.~(\ref{eq:smalleps}),
a rotation by $\varphi$ is equivalent to a shift in $\xi(t)$, one finds
in general
\begin{eqnarray}
\hat u_1(t) &=& -\pi_\rmn{E}\,\rho(t) \sin\beta\,\cos \zeta(t)\,, \nonumber \\
\hat u_2(t) &=& \pi_\rmn{E}\,\rho(t) \sin\zeta(t)\,,
\label{eq:Earthparallax}
\end{eqnarray}
where $\zeta(t) = \xi(t)-\varphi$, and $\xi(t)$ being defined
by Eq.~(\ref{eq:xi}).

If we adopt the orientation angle $\psi$ of the source trajectory,
given by $\vec u_\rmn{LS}(t)$, Eq.~(\ref{eq:sourcemotion}),
as referring to the $(\hat u_1,\hat u_2)$ coordinate axes, we find
for the total motion $\hat {\vec u}(t) = \vec u_\rmn{LS}(t) + \hat {\vec u}
(t)$, so that its absolute square reads
\begin{eqnarray}
[u(t)]^2 & = & u_0^2 + \omega_\rmn{E}^2\,(t-t_0)^2\,+\nonumber \\
& & \hspace*{-3em} +\,2\,\pi_\rmn{E}\,\rho(t)\left\{
\sin \zeta(t)\left[u_0\,\cos \psi+\omega_\rmn{E}\,(t-t_0)\,
\sin\psi\right]\right. \,+\nonumber \\
& & \hspace*{-3em} \quad +\left.\,\sin\beta\,\cos\zeta(t)\left[u_0\,
\sin\psi
-\omega_\rmn{E}\,(t-t_0)\,\cos\psi\right]\right\}\,+ \nonumber \\
& & \hspace*{-3em} +\,\pi_\rmn{E}^2\,[\rho(t)]^2\,\left(
\sin^2 \zeta(t)+ \sin^2 \beta\,\cos^2 \zeta(t)\right)\,.
\label{eq:usqparallax}
\end{eqnarray}
With the Earth's orbit defined with respect to the source at
ecliptical coordinates $(\beta,\lambda)$, only the
microlensing parallax $\pi_\mathrm{E} > 0$, determining the strength of the
parallax effect, and the angle $\psi$, defining the orientation of the source trajectory, are free parameters along with $(u_0,t_0,\omega_\rmn{E},\rho_\star)$ that define the magnification corresponding to the ordinary light curve including finite-source effects.
Since, with this parametrization, $u_0$ refers to the minimal impact
of the heliocentric trajectory, $t_0$ no longer coincides with the
maximum magnification.

\subsection{Stellar reflex motion due to orbiting planet}
\label{sec:reflex}
Given that a planet and its host star orbit their common barycentre
as described by Eq.~(\ref{eq:barymotion}), and that the motion of the observer and the source star are equivalent, it is obvious that the
periodic displacement of the observed source star due to the orbiting planet and the annual parallax due to the revolution of the Earth around
the Sun take {\em exactly} the same form.

In analogy to the discussion of the previous subsection, let us
define a parameter
\begin{equation}
\chi_\rmn{E} = \frac{m_\rmn{p}}{m}\,\frac{a}{D_\mathrm{S}\,\theta_\rmn{E}}
> 0\,,
\label{eq:defkappa}
\end{equation}
which takes over the role held by $\pi_\rmn{E}$ in measuring the strength of the effect on the microlensing light curve.
Let $i$ denote the inclination of the orbit with respect to a plane
perpendicular to the line-of-sight, and let $\varphi$ denote a longitude
in the orbital plane that decreases with the motion, where $\varphi = 0$ at periastron. The position of the source star is then given by the
orbital-plane coordinates $(i,\varphi)$, and similar to before, one finds
with $(\delta u)_\rmn{S} = (\delta r)_\rmn{S}/(D_\rmn{S}\,\theta_\rmn{E})$
\begin{equation}
\left(\begin{array}{c}
\hat u_1(t) \\ \hat u_2(t)
\end{array}\right)
= - \frac{\chi_\rmn{E}}{a}\,{\cal R}(i,\varphi)\,
\left(\begin{array}{c}
x(t) \\ y(t)
\end{array}\right) \,,
\end{equation}
with ${\cal R}$ given by Eq.~(\ref{eq:rotmatrix}), where the additional
sign results from considering the motion of the source star
rather than its planet.

More explicitly, for small eccentricity $\varepsilon$, one obtains
with Eq.~(\ref{eq:smalleps})
\begin{eqnarray}
\hat u_1(t) &=& \chi_\rmn{E}\,\rho(t) \sin i\,\cos \zeta(t)\,, \nonumber \\
\hat u_2(t) &=& -\chi_\rmn{E}\,\rho(t) \sin\zeta(t)\,,
\end{eqnarray}
with $\zeta(t) = \xi(t)-\varphi$ and $\xi(t)$ being defined
by Eq.~(\ref{eq:xi}). Let us choose the coordinate axes of
the motion of the barycentre $\vec u_\rmn{LS}(t)$, as given by Eq.~(\ref{eq:sourcemotion}), as those of the $(\hat u_1, \hat u_2)$
coordinate system. This yields a total motion
 $\vec u(t) = \vec u_\rmn{LS}(t)+ \hat {\vec u}(t)$, whose absolute square is given by
\begin{eqnarray}
[u(t)]^2 & = & u_0^2 + \omega_\rmn{E}^2\,(t-t_0)^2\,+\nonumber \\
& & \hspace*{-3em} -\,2\,\chi_\rmn{E}\,\rho(t)\left\{
\sin \zeta(t)\left[u_0\,\cos \psi+\omega_\rmn{E}\,(t-t_0)\,
\sin\psi\right]\right. \,+\nonumber \\
& & \hspace*{-3em} \quad +\left.\,\sin i\,\cos\zeta(t)\left[u_0\,
\sin\psi
-\omega_\rmn{E}\,(t-t_0)\,\cos\psi\right]\right\}\,+ \nonumber \\
& & \hspace*{-3em} +\,\chi_\rmn{E}^2\,[\rho(t)]^2\,\left(
\sin^2 \zeta(t)+ \sin^2 i\,\cos^2 \zeta(t)\right)\,.
\label{eq:usqsourcemot}
\end{eqnarray}
One indeed realizes that Eqs.~(\ref{eq:usqparallax}) and~(\ref{eq:usqsourcemot}) have an identical form, where only
$\pi_\mathrm{E} \leftrightarrow -\chi_\rmn{E}$ and $\beta \leftrightarrow i$.
Correspondingly, $u_0$ refers to the minimal impact of the barycentre, so that again the epoch $t_0$ does in general not mark a magnification maximum.

However, in contrast to the effect
of annual parallax, in addition to the strength parameter $\chi_\rmn{E}$ and the direction of source (centre-of-mass) motion,
characterized by the angle $\psi$, there are a further 5 free parameters,
namely the orbital eccentricity $\varepsilon$,
the orbital frequency $\Omega$, the orbital inclination $i$, the orbital longitude $\varphi$, and the time of periastron $t_\rmn{p}$.
In principle, the detection of reflex motion of the source star can be
confused with the motion of the Earth around the Sun (the sign between
$\pi_\rmn{E}$ and $\chi_\rmn{E}$ can simply be accounted for by
$\psi \leftrightarrow \psi+\upi$), and we should take
extreme care in avoiding to re-detect the habitable planet that we
ourselves live on. It therefore needs to be ensured that at least one
of the parameters determining the orbital motion is found to be
incompatible with annual parallax.

For circular orbits, i.e.\ $\varepsilon = 0$, the time of periastron
$t_\rmn{p}$ becomes arbitrary, and we can choose $t_0$ as reference
epoch instead, so that along with $\rho(t) \equiv 1$, one can define
$\zeta(t) = \Omega\,(t-t_0)-\varphi$. In this case, besides $(u_0,t_0,\omega_\rmn{E},\rho_\star)$, which
describe the ordinary symmetric finite-source light curve in the absence of planets, the total number of free parameters reduces
to 5, namely $(\chi_\rmn{E}, \psi, \Omega, i, \varphi)$.

Another special case arises if the orbital plane is perpendicular to
the line-of-sight, i.e.\ $i= \pm \frac{\upi}{2}$. If this happens,
modifying the angles $\varphi$ or $\psi$ by the same amount has
an identical effect, so that only one of them can be included in
a set of independent free parameters, whereas the other is obsolete.

\section{Prospects for planet detection}
\label{sec:detect}

\subsection{Strength of effect and favoured scenarios}

For a rough assessment of the amplitude of perturbations caused by
the orbital source motion, and for an identification of the favoured scenarios, as well as on dependencies on system parameters, let us
first have a look at the strength parameter $\chi_\rmn{E} = (m_\rmn{p}/M)\,
[a/(D_\rmn{S}\,\theta_\rmn{E})]$, introduced in Eq.~(\ref{eq:defkappa}).
At the source distance $D_\rmn{S}$, the angular Einstein radius $\theta_\rmn{E}$ corresponds to a physical size
\begin{eqnarray}
D_\rmn{S}\,\theta_\rmn{E} & = & \sqrt{\frac{4GM}{c^2}\,(D_\rmn{S}-D_\rmn{L})\,
\frac{D_\rmn{S}}{D_\rmn{L}}} \nonumber \\
& & \hspace*{-5em}  = \; 2.8~\mbox{AU}\,\left(\frac{M}{1~M_{\sun}}\right)^{1/2}\,
\left(\frac{D_\rmn{S}-D_\rmn{L}}{1~\mbox{kpc}}\right)^{1/2}\,
\left(\frac{D_\rmn{L}}{D_\rmn{S}}\right)^{-1/2}\,.
\end{eqnarray}
In accordance to what was found by \citet{ham06} and \citet{sum06},
the strongest effects are therefore expected for small
$D_\rmn{S} - D_\rmn{L}$, i.e.\ the lens stars being close to the source
stars, in contrast to deviations by annual parallax, which are the
strongest for lens stars close to the observer. This means that
detections on stars in the Galactic bulge are dominated by lensing
events caused by bulge stars rather than disk stars. A small
$D_\rmn{S} - D_\rmn{L}$ also implies $D_\rmn{L}/D_\rmn{S} \sim 1$,
so that $D_\rmn{S}\,\theta_\rmn{E}$ practically depends on
the distance difference $D_\rmn{S} - D_\rmn{L}$ only.
Kepler's third law, Eq.~(\ref{eq:kepler3})
allows us to eliminate the major semi-axis $a$ in favour of the
orbital period $P$, so that
\begin{eqnarray}
\chi_\rmn{E} & = & 6.4\times 10^{-4}\,\frac{m_\rmn{p}}{M_\rmn{jup}}\,
\left(\frac{m}{1~M_{\sun}}\right)^{-2/3}\,
\left(\frac{M}{0.3~M_{\sun}}\right)^{-1/2}\,
\times \nonumber \\
& & \hspace*{-3em} \times\,
\left(\frac{P}{1~\mbox{yr}}\right)^{2/3}\,
\left(\frac{D_\rmn{S}-D_\rmn{L}}{1~\mbox{kpc}}\right)^{-1/2}\,
\left(\frac{D_\rmn{L}}{D_\rmn{S}}\right)^{1/2}\,.
\label{eq:kappaofP}
\end{eqnarray}
Please note that $M$ is the mass of the lens star, whereas $m = m_\star+m_\rmn{p}$ is the total mass of the source system.
While microlensing events prefer $M \sim 0.3~M_{\sun}$, solar-mass
source stars are much brighter, so that
those are the more reasonable target despite the fact that the mass
ratio $m_\rmn{p}/m$ for a given planet mass is smaller than for low-mass
stars. Moreover, solar-mass stars are far more likely to host gas-giant planets than low-mass stars.

For the event time-scale $t_\rmn{E}$, one finds
\begin{eqnarray}
t_\rmn{E} & = & 17\,\mbox{d}\,
\left(\frac{M}{0.3~M_{\sun}}\right)^{-1/2}\,
\left(\frac{D_\rmn{S}-D_\rmn{L}}{1~\mbox{kpc}}\right)^{1/2}\,
\left(\frac{D_\rmn{L}}{D_\rmn{S}}\right)^{1/2}\,\times \nonumber \\
& & \hspace*{-3em} \times\,
 \left(\frac{D_\rmn{L}\,\mu}{160~\mbox{km}\,\mbox{s}^{-1}}
\right)^{-1}\,,
\label{eq:timescale}
\end{eqnarray}
while for Galactic bulge-bulge lensing, $D_\rmn{L}\,\mu \sim 160~\mbox{km}\,\mbox{s}^{-1}$ and
$D_\rmn{S} - D_\rmn{L} \sim 1~\mbox{kpc}$. Therefore,
a characteristic value for the favourite scenario is given by
$t_\rmn{E} \sim 17~\mbox{d}$.

\subsection{Planetary signal}

\begin{figure}
\begin{center}
\includegraphics[width=84mm]{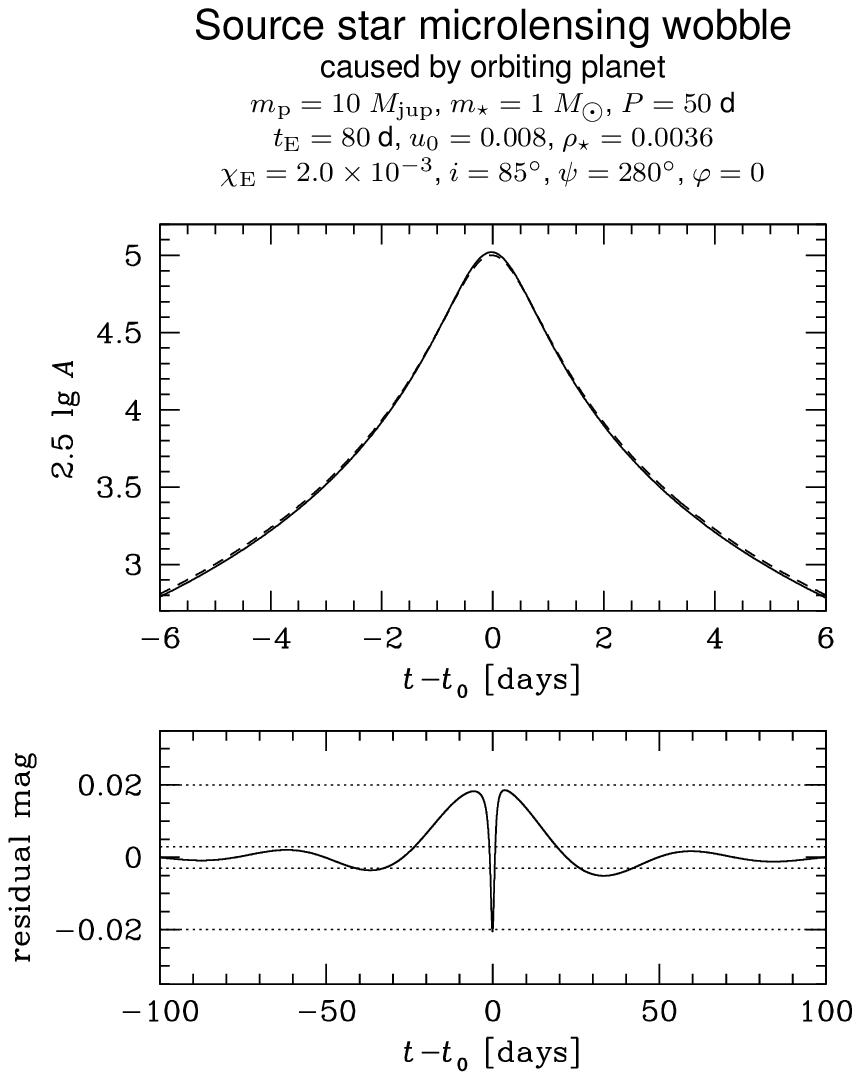}
\caption{A model light curve for which the orbital period $P$ is smaller than the event time-scale
$t_\rmn{E} \equiv \omega_\rmn{E}^{-1}$. The adopted parameters are compatible with a planet
of mass $m_\rmn{p} = 10~M_\rmn{jup}$ orbiting a star of solar mass
and solar radius at distance $D_\rmn{S} = 8.5~\mbox{kpc}$, whereas
the lens star of mass $M = 0.4~M_{\odot}$ is located at
$D_\rmn{L} = 8.0~\mbox{kpc}$ from the observer. The upper panel
shows the resulting light curve (solid) as well as a best-fitting
approximation (dashed) with an ordinary model $(u_0,t_0,\omega_\rmn{E},\rho_\star)$. The difference between these two curves is plotted in the lower panel. 
Dotted lines indicate deviations by 2 per cent or 0.3 per cent, respectively.
} \label{fig:wobble1}
\end{center}
\end{figure}
\begin{figure}
\begin{center}
\includegraphics[width=84mm]{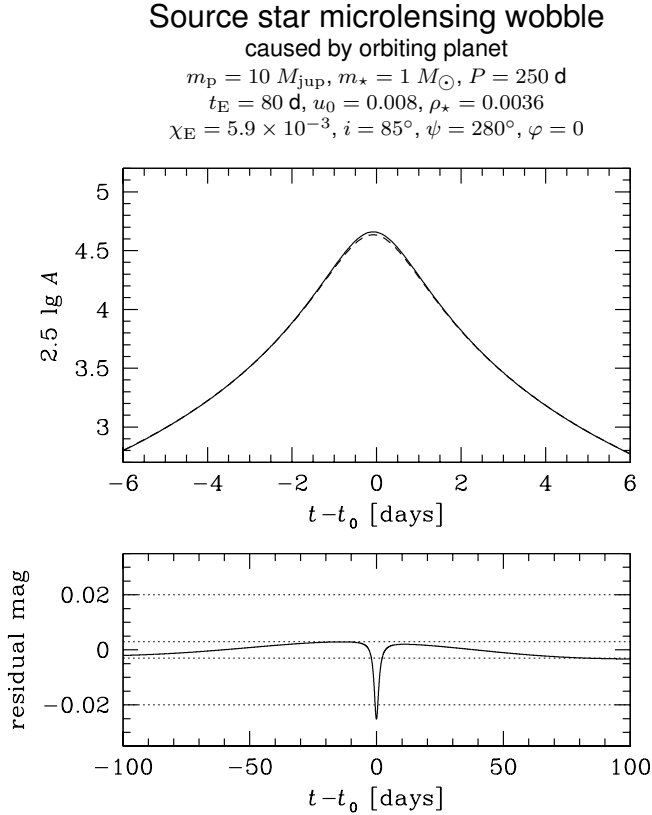}
\caption{A model light curve for which the orbital period $P$ is larger than the event time-scale
$t_\rmn{E}$. The adopted parameters are compatible with the same
scenario as chosen for Fig.~\protect\ref{fig:wobble1}. As before,
the upper panel show the resulting light curve (solid) together with a best-fitting
ordinary light curve (dashed), while the lower panel displays their difference. Dotted lines refer to deviations by 2 per cent or 0.3 per cent, respectively.} \label{fig:wobble2}
\end{center}
\end{figure}

Since displacements of the source star cause larger changes to its 
magnification the smaller the source-lens separation (and therefore, the
larger the magnification), the effects of the orbiting planet on the light curve increase with the source magnification, and the planet needs to 
be identified while the latter is substantial. Therefore,
the characteristics of the planetary signal depend on 
whether the orbital period $P$ is smaller or larger than the event
time-scale $t_\rmn{E}$, as illustrated in Figs.~\ref{fig:wobble1}
and~\ref{fig:wobble2}. As by the definition of the parameters, $u_0$ does not refer to the closest
approach between lens and source star, so that the peak magnification
differs from $A_0 \simeq u_0^{-1} = 125$, where the discrepancy is larger
for the long-period case with the larger $\chi_\rmn{E}$. Given that 
we only altered the orbital period between the two cases shown, while
leaving all other physical properties unchanged, the signal strength
parameter $\chi_\rmn{E}$ increases with the orbital period $P$,
according to Eqs.~(\ref{eq:defkappa}) and~(\ref{eq:kappaofP}).

For the case $P \ll t_\rmn{E}$, the periodic pull of the  planet
on its host star leads to detectable ripples on the observed light curve,
while the long orbital period $P$ deprives us of such a characteristic
signature for $P \gg t_\rmn{E}$. Without a good indication of $P$,
a distinction with the effect of annual parallax becomes difficult,
and to the lowest order, one only observes an acceleration effect
\citep{smith02}. However, regardless of $P/t_\rmn{E}$, a best-fitting ordinary light curve leaves us   
with a mismatch near the tip of the light curve that is not overcome by
adopting a different finite-source parameter $\rho_\star$, and allows
a detection if the impact parameter $u_0$ is sufficiently small for such
a signal to be prominent enough.

\subsection{Parameters of simulation}
\label{montcarlo}
After having identified the basic scenarios, we carried out Monte-Carlo
simulations in order to study the detectability of planets orbiting
Galactic bulge stars as a function
of various parameters that describe the lens star, the source star, and
its orbiting planet. For simplicity, we have assumed circular orbits
($\varepsilon = 0$), so that the source magnification $A[u(t);\rho_\star]$ is
described by the 9 parameters $(u_0, t_0, \omega_\rmn{E}, \rho_\star, \chi_\rmn{E},
\psi,\Omega,i,\varphi)$.

Without loss of generality, we set $t_0 = 0$. Moreover, 
we consider 
planets of mass $m_\rmn{p}= 1~M_\rmn{jup}$ or $m_\rmn{p} = 10~M_\rmn{jup}$
orbiting a source star of $m_\star = 1~M_{\sun}$.
With these choices, according
to Eq.~(\ref{eq:kappaofP}), the strength parameter $\chi_\rmn{E}$ then
becomes a function of the lens and source distances $D_\rmn{L}$ and 
$D_\rmn{S}$, for $D_\rmn{S} - D_\rmn{L} \ll D_\rmn{L}$ essentially
of $D_\rmn{S} - D_\rmn{L}$, the orbital frequency $\Omega = (2\,\upi)/P$, where $P$ denotes the orbital period, and the lens mass $M$.
In relation to the sampling interval and the event time-scales, we
generate a uniform distribution in $\lg [P/(1~\mbox{d})]$ ranging between
$3~\mbox{d}$ and $1~\mbox{yr}$. While we adopt a 'natural' uniform
distribution of impact parameters $u_0\in[0,1]$, not taking into account
any selection bias by the experiment, the phase angle $\varphi$, orientation angle $\psi$, and inclination angle $i$ are all naturally uniformly distributed, where $\varphi,\psi\in[0,2\pi)$ and $i\in[0,\pi/2]$.

The event time-scale $t_\rmn{E}$ follows from drawing a lens distance
$D_\rmn{L}$, source distance $D_\rmn{S}$, velocity $v$, and
lens mass $M$ from the adopted
distributions for the Galactic bulge described in Appendix~\ref{app:bulge}.
Moreover, with assuming a source radius $R_\star = 1~R_{\sun}$, one
obtains the source size parameter $\rho_\star = [R_*/(v\,t_\rmn{E})]\,(D_\rmn{L}/D_\rmn{S})$.






In contrast to the discussion by \citet{rah03} of the observability of parallax effects towards the Large Magellanic Cloud, which used parameters specific to the 
EROS (Experience de la Recherche d'Objets Sombres) campaign, we adopt
the simple pragmatic approach of assuming a constant photometric uncertainty for equally-spaced observations over the course of the 
microlensing event without any loss due to bad weather.
In particular, we choose (a) a sampling interval of 2~hours with
2 per cent accuracy, resembling current follow-up observations, or
(b) a sampling interval of 15~minutes with 0.3 per cent, resembling upcoming
campaigns. More precisely, we demand the fractional uncertainty of the
magnification to match the quoted value, or equivalently, the flux
after subtraction of the background to be measured that precisely.
In fact, observing campaigns need to account for such a requirement, 
or time is being wasted on taking data on strongly blended targets
without a chance to extract meaningful results.
The simple choices allow us to focus on the primary dependencies without being bound to
the variety of different existing or possible setups, which all show
different characteristics with respect to the crowding of targets,
the distribution of event impact parameters for main-sequence stars, and
the actually achieved photometry, where all these effects are correlated
with each other.

\subsection{Detectability in simulated events}

For each of the created synthetic light curves, we obtain best-fitting model parameters by means of $\chi^2$-minimization, which corresponds to a maximum-likelihood
estimate for Gaussian error bars,
for an ordinary light curve $(u_0,t_0,\omega_\mathrm{E})$ and 
independently the 8 parameters $(u_0,t_0,\omega_\mathrm{E},
\chi_\rmn{E},\psi,\Omega,i,\varphi)$ that include the description of the 
motion of the source induced by a planet in a circular orbit.
In both cases, we also determine effective best-fitting source and background fluxes $F_\rmn{S}$ and $F_\rmn{B}$, taking into account
a potential difference between the best-fitting magnification and
the true magnification arising from the simulation.
We explicitly adopt a finite angular radius $\theta_\star$ of the source 
star, but do not refit for the respective parameter $\rho_\star \equiv
\theta_\star/\theta_\rmn{E}$ in
order to save computing time and to avoid parameter degeneracies if
finite-source effects are not prominent.

A quantitative measure for the detectability of a planetary signal
then results from a likelihood-ratio test, involving the respective
$\chi^2$ minima, namely $(\chi^2_\rmn{min})^{(0)}$ and $(\chi^2_\rmn{min})^{\rmn{planet}}$. In fact, with $\Delta \chi^2_\rmn{p} \equiv
(\chi^2_\rmn{min})^{(0)}-(\chi^2_\rmn{min})^{\rmn{planet}}$ following
a $\chi^2$-distribution with 5 degrees of freedom, one finds 
a probability $P(\Delta \chi^2_\rmn{p} \geq 11.07) = 0.05$ for such a difference
to arise. At this significance level, we therefore reject the hypothesis
that an ordinary light curve explains the data whenever 
$\Delta \chi^2_\rmn{p} \geq 11.07$, and claim the detection of a signal.

\begin{figure*}
\begin{center}
\includegraphics[width=82mm]{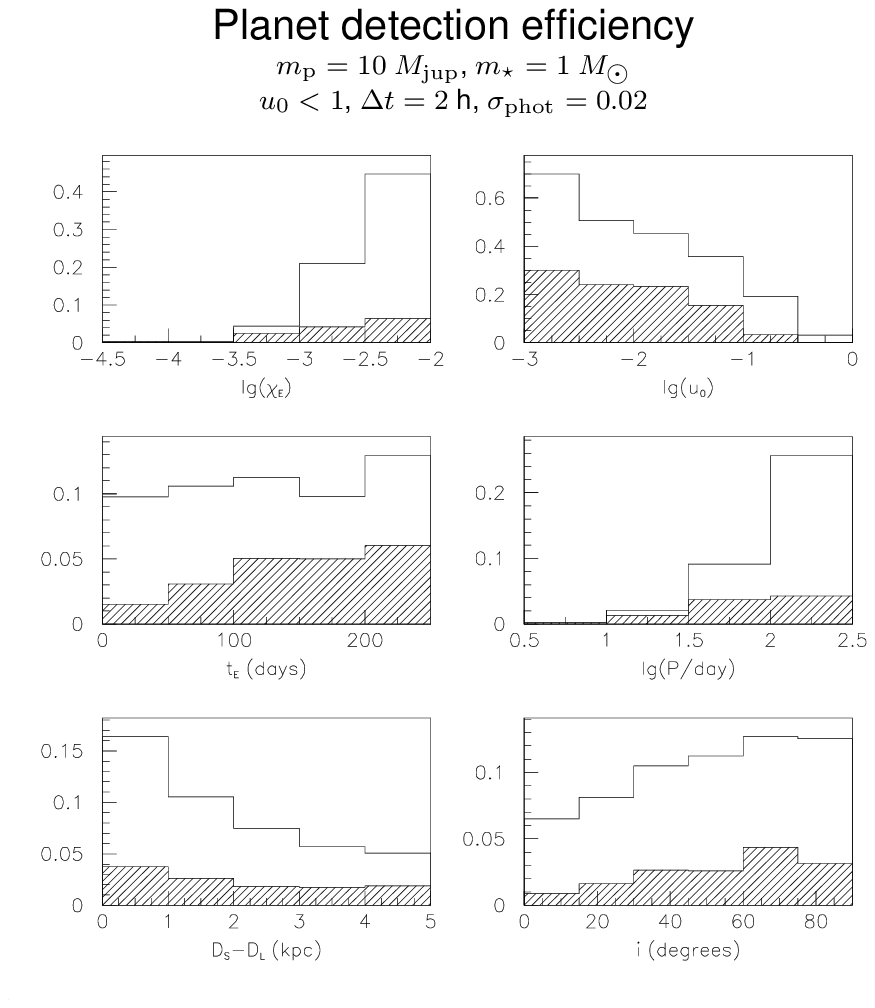}
\hfill
\includegraphics[width=82mm]{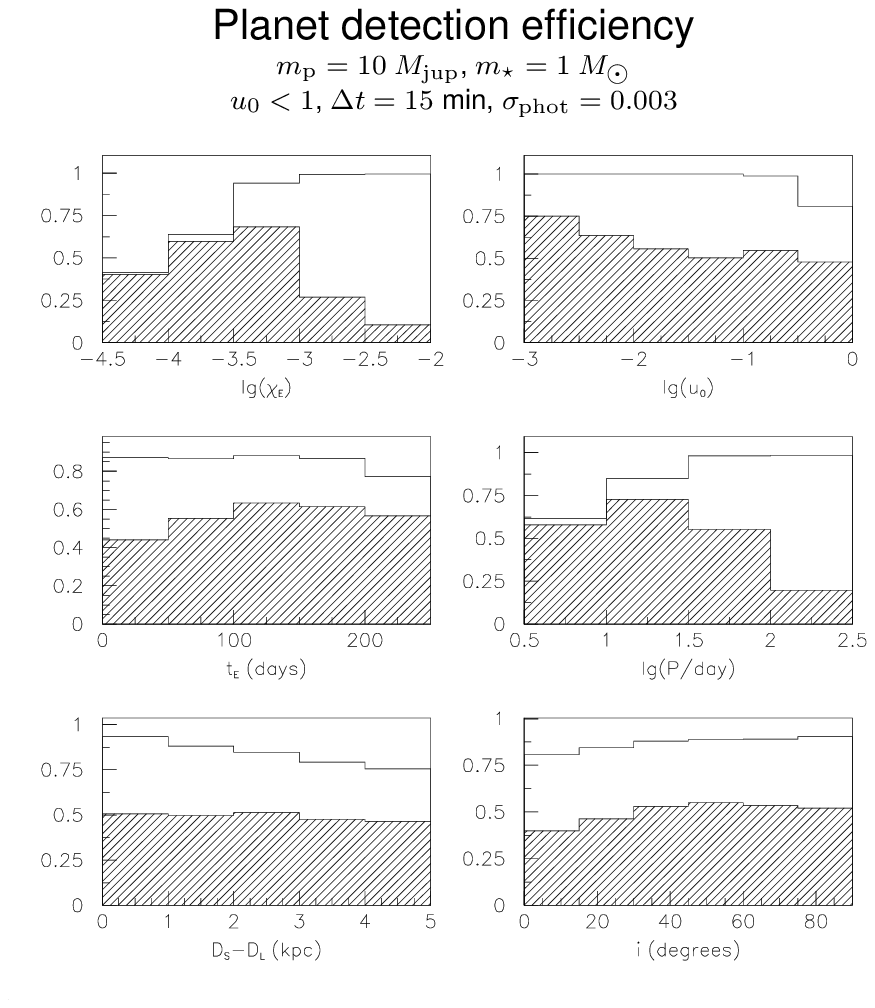}
\caption{Efficiency of detecting a planet of mass $m_\rmn{p}=10~M_\rmn{jup}$
by means of the effect of the orbital motion of its observed host star 
in the Galactic bulge with $m_\star = 1~M_{\sun}$ on the light curve 
arising from its light being bent by an intervening foreground 
Galactic bulge lens star as a function of the strength parameter $\chi_\rmn{E}$, defined by Eq.~(\ref{eq:defkappa}), the event impact parameter $u_0
\in [0,1]$,
the event time-scale $t_\rmn{E}$, the orbital period $P$, the lens-source
distance $D_\rmn{S}-D_\rmn{L}$, and the orbital inclination $i$. For the left panel, we assume
observations with 2 per cent photometric accuracy regularly spaced at 2~h intervals, while the results shown in the right panel correspond to increased observational capabilities of 0.3 per cent
photometric accuracy and a sampling interval of 15~min. The hatched area refers to cases where the signal is characteristic enough to be distinguished from false positives.}
\label{bulge10} 
\end{center}
\end{figure*}

\begin{figure*}
\begin{center}
\includegraphics[width=82mm]{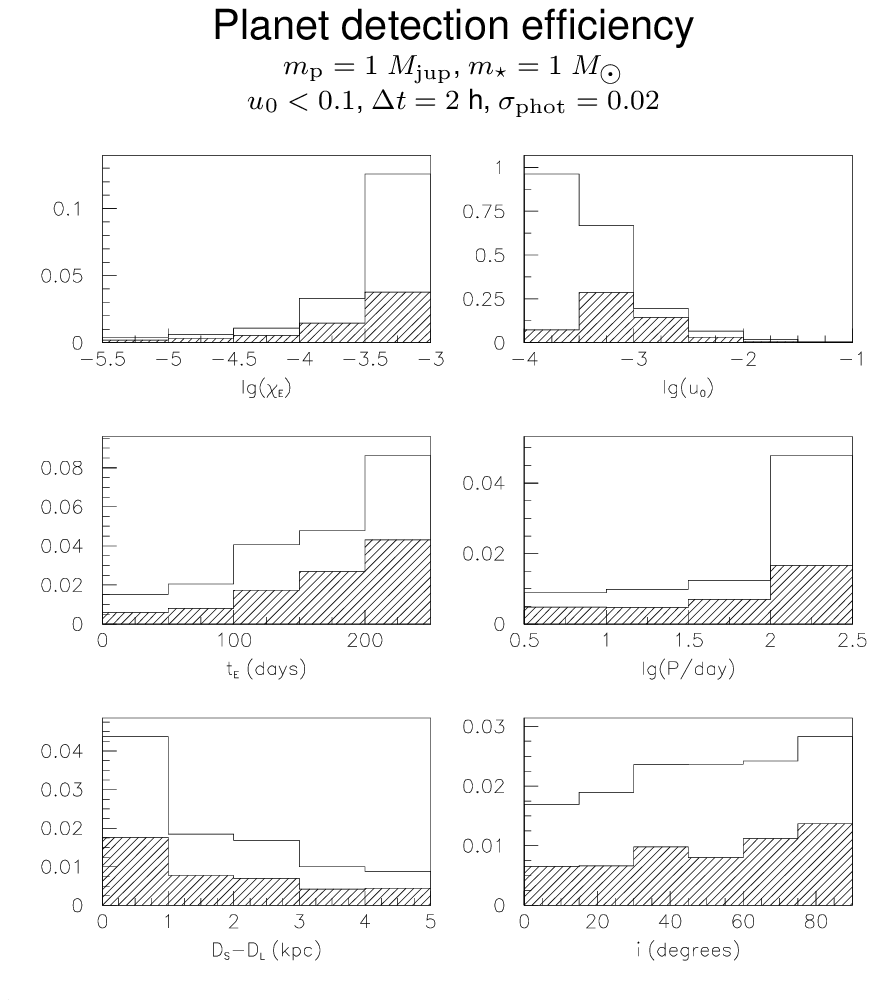}
\hfill
\includegraphics[width=82mm]{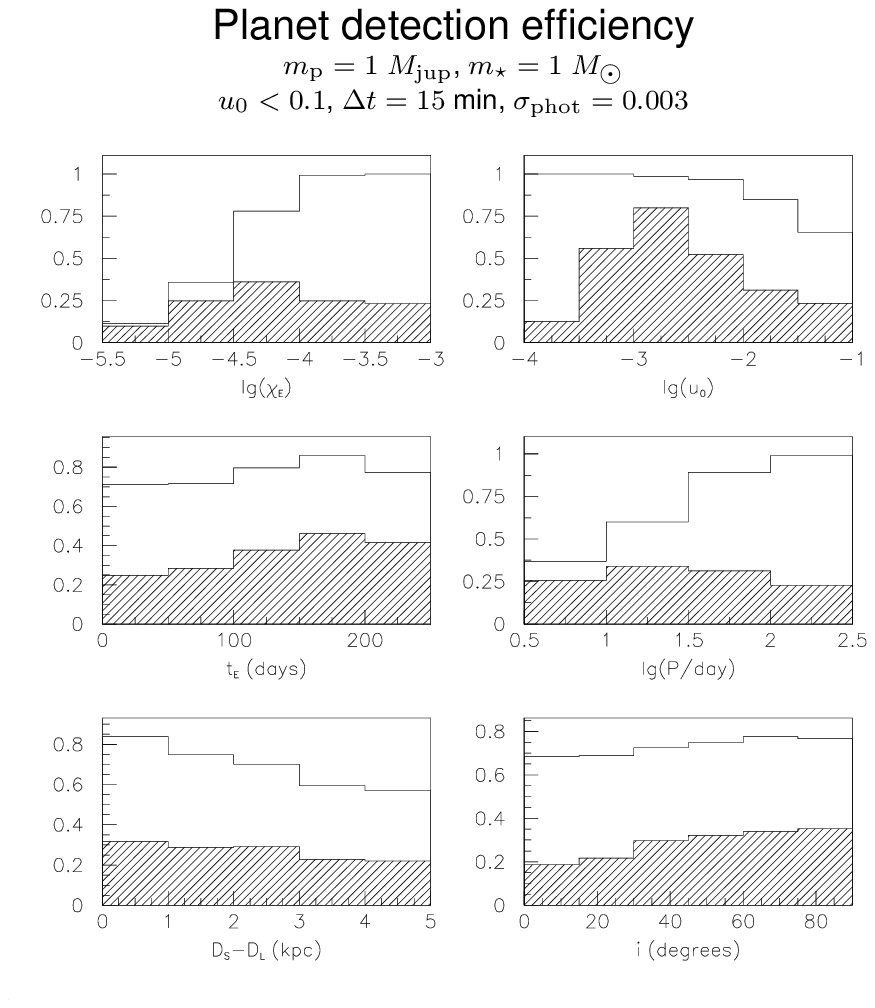}
\caption{Detection efficiency for a Jupiter-mass planet ($m_\rmn{p}=1~M_\rmn{jup}$) from the influence of source-star orbital motion on the
microlensing light curve for two different observational capabilities,
similar to Fig.~\ref{bulge10}, but considering only events with smaller impact parameters $u_0 \in [0,0.1]$.}
\label{bulge1} 
\end{center}
\end{figure*}

A substantial fraction of such 'detections' however do not involve characteristic
features, so that instead of the assumed effect of periodic source motion
revealing the presence of an orbiting planet,
these could be of different origin if found in an observed event.
In order to account for false positives, we therefore adopt a second
criterion. In analogy to the primary criterion, we determine the
5 parameters $(u_0,t_0,\omega_\mathrm{E}, \pi_\rmn{E}, \psi)$ that include the 
annual parallax due to the Earth's revolution, and carry out a similar
likelihood-ratio test based on the criterion $\Delta \chi^2_{\scriptscriptstyle\oplus} \equiv (\chi^2_\rmn{min})^{(0)}-(\chi^2_\rmn{min})^{\rmn{Earth}} \geq 5.99$, where the adopted
value results from $P(\Delta \chi^2_{\scriptscriptstyle\oplus} \geq 5.99) = 0.05$ for
a $\chi^2$-distribution with 2 degrees of freedom.
We then consider the planetary signal to be {\em characteristic}
if the detection of a signal is significant ($\Delta \chi^2_\rmn{p} \geq 11.07$), while a signature of annual parallax
 is not ($\Delta \chi^2_{\scriptscriptstyle\oplus} < 5.99$); and the signal to be featureless otherwise.
With this procedure, we not only reject events with deviations that are
likely to be caused by the motion of the Earth, but in an elegant way,
we also get rid of all uncharacteristic deviations that are compatible,
which could e.g.\ be due to differences in the optimal finite-source parameter or weak effects resulting from binarity of the lens or source
object. At the end, we are left only with cases for which either the 
observed deviation is characteristic for the periodic motion of the
source star, or where an asymmetric feature is closely mimicked by effects
other than the annual parallax.



The relative abundances of planet detections amongst all simulated
events as a function of various parameters, namely the event
impact parameter $u_0$, the orbital period $P = (2\,\upi)/\Omega$, the
event time-scale $t_\rmn{E} \equiv \omega_\rmn{E}^{-1}$, the 
lens-source distance $D_\rmn{S}-D_\rmn{L}$, the orbital inclination $i$,
and the resulting strength parameter $\chi_\rmn{E}$ are shown in Figs.~\ref{bulge10} and~\ref{bulge1} corresponding to planets of mass $m_\rmn{p} = 10~M_\rmn{jup}$ or
$m_\rmn{p} = 1~M_\rmn{jup}$, respectively, orbiting solar-mass Galactic
bulge stars, and the two adopted observational capabilities with regard
to the sampling frequency and achievable photometric accuracy.

For $m_\rmn{p} = 10~M_\rmn{jup}$ and the less favourable 2 per cent
accuracy and 2~h sampling, about 1/4 of the significant deviations are characteristic for a detection of orbital stellar motion against false positives, providing a detection efficiency of $\sim\,$2.5 per cent over
the adopted sample. Small impact parameters strongly support the detection,
with efficiencies reaching $\sim\,24$ per cent for characteristic signals arising amongst the smaller number of events
with $u_0 \la 0.02$ ($A_0 \ga 50$). While nearly all detected signals 
are uncharacteristic for $u_0 \ga 0.1$, characteristic and uncharacteristic
detections are of similar frequency for smaller $u_0$. The detection efficiency is found to
increase towards larger event time-scale $t_\rmn{E}$ and orbital period $P$,
as well as to favour smaller $D_\rmn{S}-D_\rmn{L}$ and face-on
orbits $i \sim 90^{\circ}$ over edge-on orbits $i \sim 0^{\circ}$.

With a close-to-continuous sampling at 15~min intervals and 0.3 per
cent photometric accuracy, planets of $10~M_{\rm jup}$ become hard to miss,
with more than 85 per cent of them showing detectable signals, and a
small $u_0$ no longer being in strong favour of a detection.
As compared to the less favourable observational capabilities, more than
half of the detected deviations contain characteristic features. 
This is a 
substantial improvement, pushing the detectability of planets by 
a factor of $\sim\,20$. 
The much larger gain on the detectability with the improved capabilities for shorter periods accounts for a decrease of the characteristic 
detections towards larger orbital periods (whereas the fraction of uncharacteristic detections increases). In particular, for all the
shorter orbital periods $P \la 10~\mbox{d}$, the planetary signal
is characteristic, and almost $60$ per cent of the planets are found.
One might be puzzled by the fact that for the largest encountered strength
parameters $\chi_\mathrm{E}$, the prospects for a
characteristic detection decrease. While it seems a bit surprising, 
this behaviour results from $\chi_\mathrm{E}$ being correlated with
the orbital period $P$, and larger orbital periods mean that the 
light curves contain less characteristic features.

For the prospects of detecting a Jupiter-mass planet, the results of
our simulations reveal mostly the same trends with the parameters as for $10~M_\rmn{jup}$. While for such less massive planets, one is extremely unlikely to succeed unless
$u_0 \la 0.03$ with a strategy of 2~h sampling at 2 per cent photometric 
accuracy, and a substantial rise of the detection efficiency towards smaller $u_0$ even for the more favourable observational setup, very small $u_0 \la 0.0003$ are not optimal either, given that
finite-source effects wash out signals in that regime. On average, characteristic detections happen for $\sim\,$0.9 per cent of all events with $u_0 < 0.1$, but for $\sim\,$12 per cent of events with $u_0 \leq 0.005$. With the adopted better observational capabilities, these values rise to $\sim\,$30 per cent or $\sim\,$70 per cent, respectively.

Given that the same value of the strength parameter $\chi_\mathrm{E}$ 
for a larger planet mass $m_\rmn{p}$ implies a smaller orbital period $P$,
which eases the detection of a planetary signal, the detection efficiency
for more massive planets at same  $\chi_\mathrm{E}$ should be larger.
However, the smaller detection efficiency for some of the ranges shown in the figures
is a result of averaging over events with $u_0 \in [0,0.1]$ for $m_\rmn{p} = 1~M_\rmn{jup}$, whereas an average over $u_0 \in [0,1]$ has been taken
for $m_\rmn{p} = 10~M_\rmn{jup}$.

\begin{figure*}
\begin{center}
\includegraphics[width=82mm]{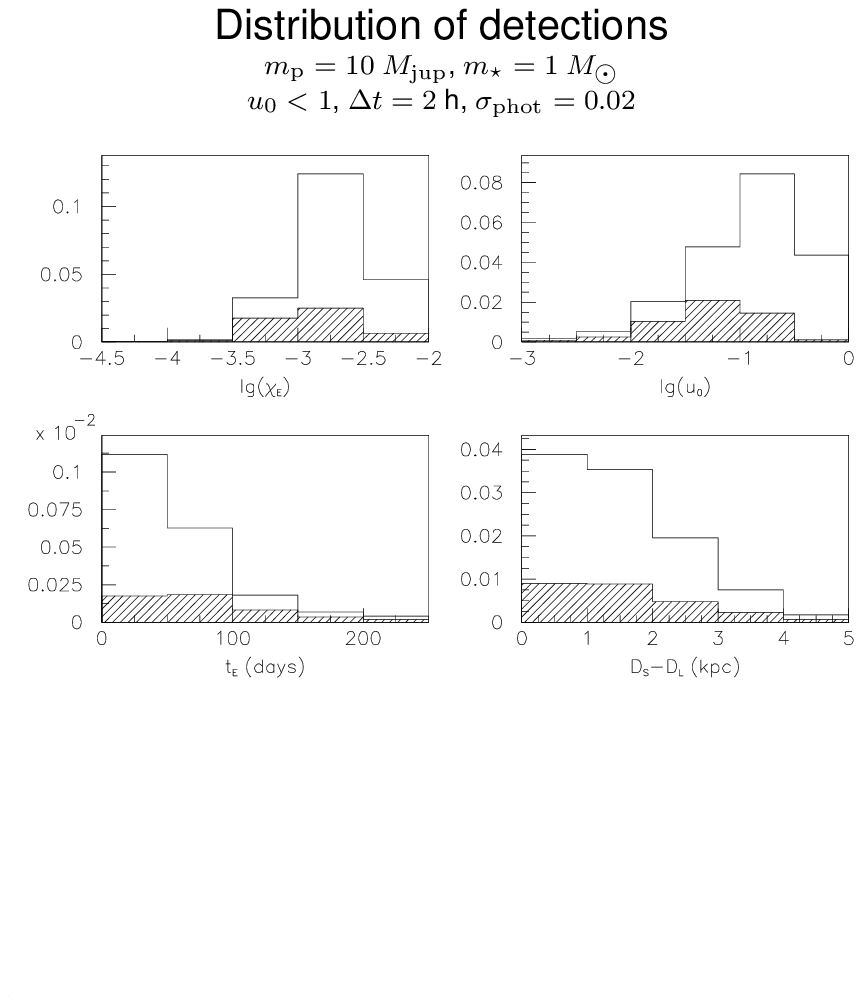}
\hfill
\includegraphics[width=82mm]{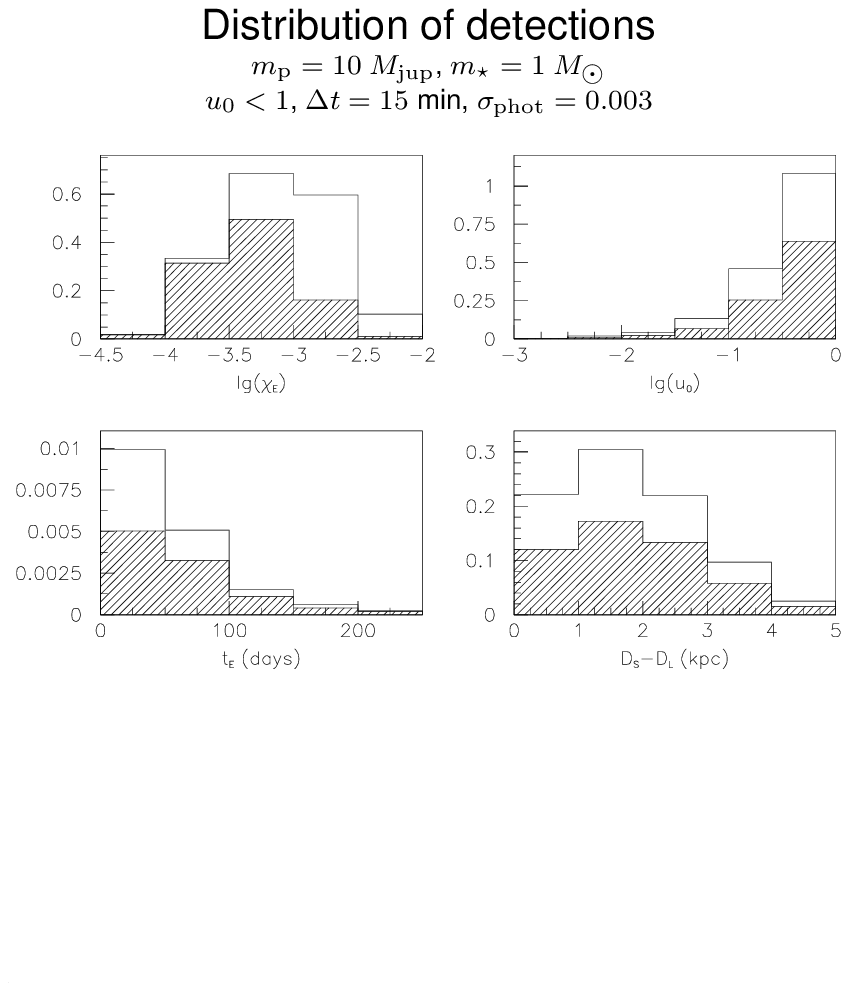}
\caption{Distribution with the signal strength parameter $\chi_\rmn{E}$,
the event impact parameter $u_0 \in [0,1]$, the event time-scale $t_\rmn{E}$, or the lens-source distance $D_\rmn{S}-D_\rmn{L}$ 
of the detection of a planet of mass
$m_\rmn{p} = 10~M_\rmn{jup}$ orbiting a Galactic bulge star with 
$m_\star = 1~M_{\sun}$ from the change in the microlensing light curve
resulting from the periodic displacement of the observed star caused by
the gravitational pull of the orbiting planet. As for Fig.~\protect\ref{bulge10}, the two panels refer to different observational capabilities, namely 2~h sampling with 2 per cent accuracy or
15~min sampling with 0.3 per cent accuracy, and the hatched area marks
the characteristic detections. The total area of the bins gives the
average detection efficiency. For the uniformly-distributed quantities
$\lg [P/(1~\mbox{yr})]$ and $i$, the distribution of the detections is proportional to the detection efficiencies plotted in 
Fig.~\protect\ref{bulge10}.}
\label{fig:dist10} 
\end{center}
\end{figure*}

\begin{figure*}
\begin{center}
\includegraphics[width=82mm]{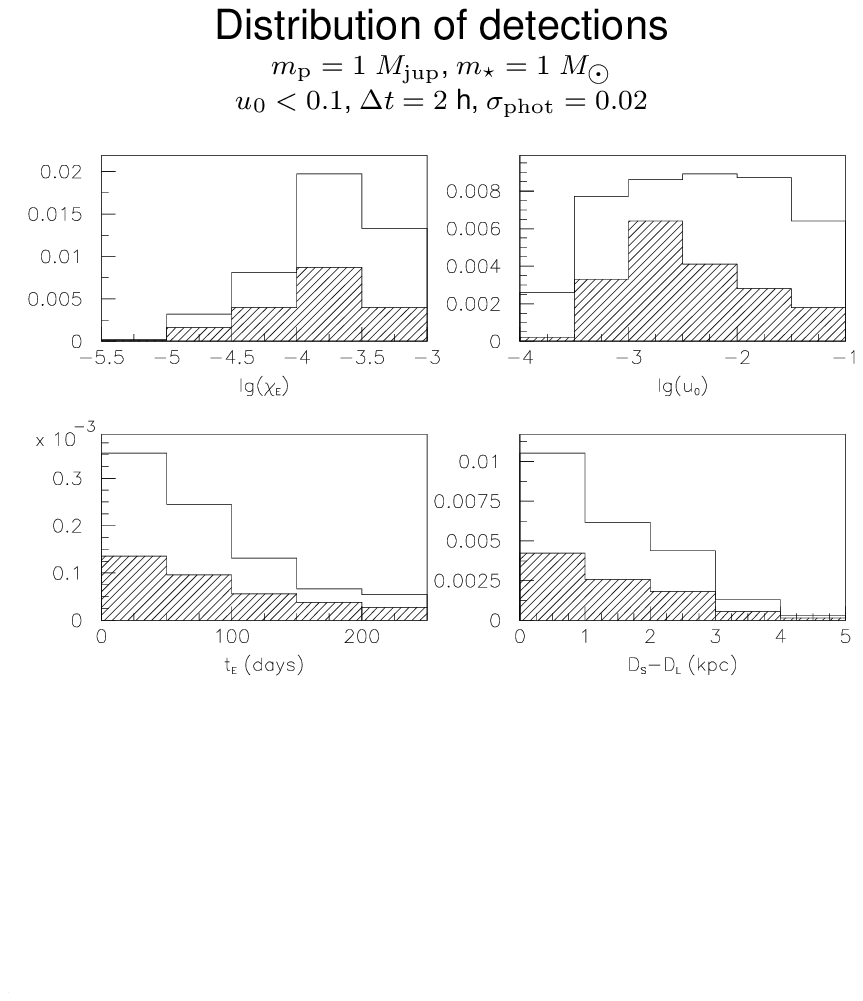}
\hfill
\includegraphics[width=82mm]{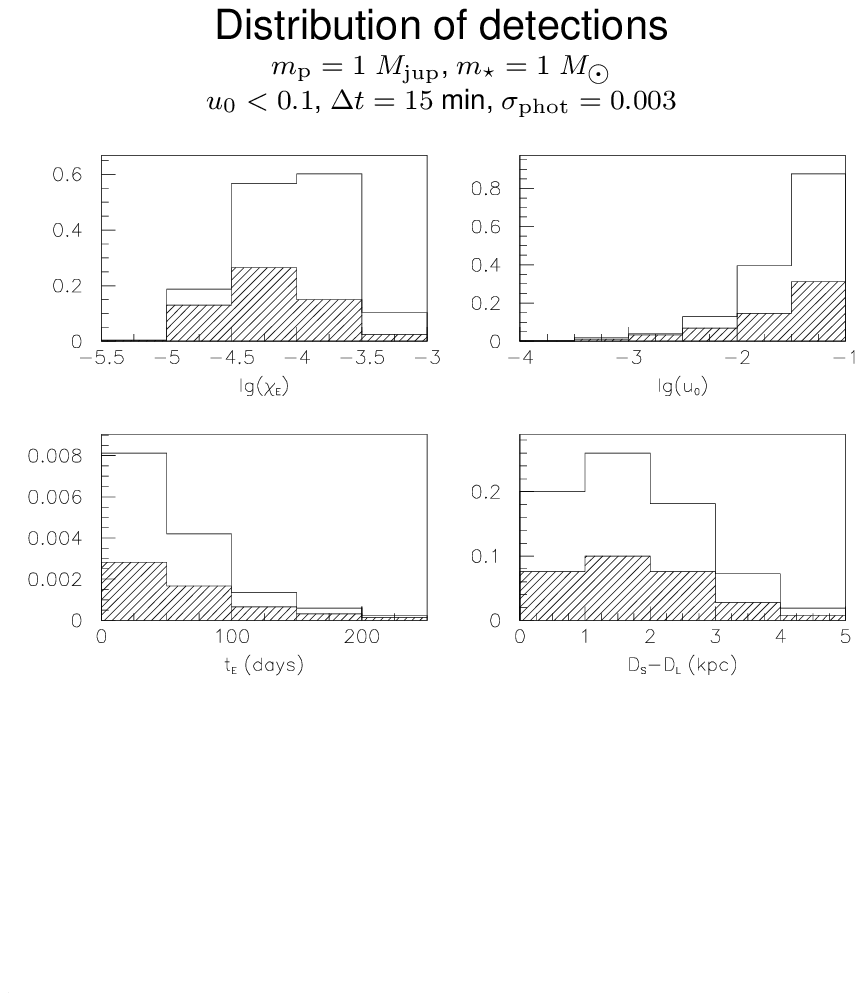}
\caption{Distribution of the planet detections with several parameters,
similar to Fig.~\ref{fig:dist10}, but now for a planet mass of $m_\rmn{p}=1~M_\rmn{jup}$ and the impact parameter in the range
$u_0 \in [0,0.1]$.}
\label{fig:dist1} 
\end{center}
\end{figure*}

Rather than the planet detection efficiencies for certain parameter ranges,
Figs,~\ref{fig:dist10} and~\ref{fig:dist1} show the distribution of the
detections with these parameters, i.e.\ in contrast to before, the 
distribution of the respective parameter ranges among the simulated events
has been considered (which makes no difference for uniformly distributed
quantities, which have been omitted). In all four considered cases (two planet masses and two sets of observational capabilities), the fact 
that microlensing events preferentially arise with 
time-scales $5~\mbox{d} \leq t_\rmn{E} \leq 80~\mbox{d}$ leaves 
such small $t_\rmn{E}$ with the largest number of 
planet detections, despite the fact the the detection efficiency 
increases with $t_\rmn{E}$.
With 2~h sampling and 2 per cent accuracy, there are different
preferences for the impact parameter $u_0$, depending on the mass
of the planet. While for $m_\rmn{p} = 10~M_\rmn{jup}$, the bulk of 
detections is expected from $0.01 \la u_0 \la 0.3$, the difficulty of revealing a signal for larger $u_0$ in the case of
for $m_\rmn{p} = 1~M_\rmn{jup}$ moves the preferred range
to $0.0003 \la u_0 \la 0.03$. For both planet masses, the better
observational capabilities of 15~min sampling and 0.3 per cent 
photometric accuracy lead to the larger number of events with larger $u_0$ 
providing the larger number of detections.

\subsection{The challenge of detecting extragalactic planets}
No other technique but gravitational microlensing has so far been
suggested to achieve the detection of planets orbiting stars in
neighbouring galaxies, such as M31, within foreseeable
time \citep{cov00,chu06}. Given that the signal strength $\chi_\rmn{E}$
related to the periodic reflex motion of the source star due to an orbiting planet lacks of practical dependence on the distance $D_\rmn{S}$, one might
hope that this planet detection channel could also be a viable 
alternative for extragalactic planets.

As compared to the Galactic bulge, there are however some severe 
difficulties that make any such attempts a quite demanding challenge.
Most importantly, one faces the problem that the target stars are not
resolved, so that the effectively achieved photometric accuracy is 
quite limited. Unless major technology leaps are made, one is 
therefore restricted
to very strong signals, which are unlikely to occur. Catching
main-sequence source stars rather than giants will only be possible
during phases of extreme magnification, limiting the number of suitable
events further. Beyond that, detections on M31 targets are disfavoured due to the larger typical
$D_\rmn{S}-D_\rmn{L} \sim 10~\mbox{kpc}$ as compared to the Galactic bulge,
where $D_\rmn{S}-D_\rmn{L} \sim 1~\mbox{kpc}$.

\subsection{Does the planet orbit the source star?}
So far, we have assumed by construction that the periodic alteration of the relative position between lens and source star as seen from Earth is the result of a planet orbiting the observed source star, and only the revolution of the Earth itself has been considered as an alternative.
However, as explicitly pointed out to us by V.~Bozza (private communication), a similar effect might also arise from a planet orbiting
the foreground lens star rather than the observed background source star.

We are quite familiar with planets orbiting the foreground lens star
revealing their presence due to them affecting the gravitational bending
of light received from the source star \citep{mao91}. Notably, the
arising characteristic signal represents a snapshot of the planet at its
current angular separation from its host star, and does not depend on
its orbital period. Moreover, the chances of detecting a planetary signal
rely on a resonance of the angular Einstein radius $\theta_\rmn{E}$ with
the angular separation of the planet $\theta_\rmn{p} = d\,\theta_\rmn{E}$,
whereas the planet is likely to escape detection for either $d \ll 1$ or
$d \gg 1$.

It is obvious that similar to a source star of mass $m$, a lens star
of mass $M$ also exhibits a periodic shift due to a planet orbiting
at a semi-major axis $a$, whose strength in full analogy to the
discussion of Sect.~\ref{sec:reflex} is given by
\begin{equation}
\lambda_\rmn{E} = \frac{m_\rmn{p}}{M}\,\frac{a}{D_\rmn{L}\,\theta_\rmn{E}}\,.
\end{equation}
Inserting characteristic values yields
\begin{eqnarray}
\lambda_\rmn{E} & = & 1.4\times 10^{-3}\,\frac{m_\rmn{p}}{M_\rmn{jup}}\,
\left(\frac{M}{0.3~M_{\sun}}\right)^{-7/6}\,
\times \nonumber \\
& & \hspace*{-3em} \times\,
\left(\frac{P}{1~\mbox{yr}}\right)^{2/3}\,
\left(\frac{D_\rmn{S}-D_\rmn{L}}{1~\mbox{kpc}}\right)^{-1/2}\,
\left(\frac{D_\rmn{L}}{D_\rmn{S}}\right)^{-1/2}\,,
\label{eq:lambdaE}
\end{eqnarray}
and by comparing this expression with the corresponding Eq.~(\ref{eq:kappaofP}), one sees that the detection as compared to planets orbiting the 
source star is not only facilitated by the smaller typical stellar mass,
but lens distances not only close to the source star, but also close to
the observer provide favourable configurations.

Assuming circular orbits, a typical orbital radius is given by
$a \sim 1.36~d\,D_\rmn{L}\,\theta_\rmn{E}$ \citep{Do:Esti2}, so that
for $d = 1$, characteristic orbital periods are of the order of
\begin{equation}
P_0 = 5.6~\mbox{yr}\,\left(\frac{D_\rmn{L}-D_\rmn{S}}{1~\mbox{kpc}}
\right)^{3/4}\,\left(\frac{D_\rmn{L}}{D_\rmn{S}}\right)^{3/4}\,
\left(\frac{M}{0.3~M_{\sun}}\right)^{1/4}\,,
\end{equation}
which usually exceed those for which we expect to detect signatures
of the period reflex motion. Given that $P_0 \propto d^{3/2}$, we find
in particular that orbital periods $P_0 \la 50~\mbox{days}$ roughly
correspond to separation parameters $d \la 0.08$, so that substantial
deviations due to gravitational bending of light are not likely to occur.
In fact, such can always be checked for by means of constructing model
light curves that take this effect into account.

However, as a consequence, we are facing the situation that even in the case that one identifies a characteristic signal that cannot arise from the revolution of the Earth, it remains
possible that one ends up with two competing interpretations putting the
planet in orbit around the source star or the lens star, respectively.

\subsection{Properties of the planet and its orbit}

For the 'usual' channel of planet detection by microlensing, the 
properties of the planet and its orbit
affect the light curve only by means of two dimensionless parameters,
which can be taken as the planet-to-star mass ratio $q$ and the
separation parameter $d$, where $d\,\theta_\rmn{E}$ is the instantaneous
angular separation of the planet from its host star. With the 
technique being most sensitive to planets around $d \sim 1$, and planetary
signals lasting between hours and days, these are substantially smaller than the orbital period $P$, and therefore the only information about the
planetary orbit is provided by the separation parameter $d$, whereas
neither the orbital eccentricity $\varepsilon$, or the orbital inclination $i$ can be inferred from the snapshot. 

The mass of the planet $M_\rmn{p} = q\,M$ is related to the mass of the
lens star $M$, which does not follow directly from the light curve,
but in general needs to be inferred probabilistically from the
event time-scale $t_\rmn{E} = \theta_\rmn{E}/\mu$, with the angular
Einstein radius, given by Eq.~(\ref{eq:thetaE}), being a function
of the lens mass as well as of the lens and source distances $D_\rmn{L}$
and $D_\rmn{S}$. As discussed in detail by \citet{Do:Esti2}, this requires
the adoption of a kinematic model of the Galaxy as well as of mass
functions of the underlying stellar populations that make up the lens
stars.
Light curves that involve planets with small masses however are likely to be influenced by the finite angular size $\theta_\star$ of the source star. 
With the possibility to determine $\theta_\star$ from stellar typing
based on its magnitude and colour, and better with a spectrum, the
extractable time-scale $t_\star = \theta_\star/\mu$, in which the source
moves by its own angular radius with respect to the lens, allows 
to infer the proper motion $\mu$, and thereby with the event time-scale
$t_\mathrm{E} = \theta_\rmn{E}/\mu$ of the angular Einstein radius $\theta_\rmn{E}$. With a reliable estimate for the source distance $D_\rmn{S}$,
the mass of the lens star $M$ thereby only becomes a function of the lens
distance $D_\rmn{L}$, reducing the uncertainty substantially. However,
the mass of the lens star $M$, and thereby the mass of the planet
$M_\rmn{p} = q\,M$ only follows from the observed light curve, if moreover
the microlensing parallax $\pi_\rmn{E} = \pi_\mathrm{LS}/\theta_\rmn{E}$
can be determined. As for the lens mass $M$, only a probability density
can in general be obtained for the instantaneous
physical projected separation $\hat r = d\,D_\rmn{L}\,\theta_\rmn{E}$,
while a stochastic distribution for the orbital semi-major axis $a$
further follows with an orbital deprojection and assumption of a 
distribution of orbital eccentricities $\varepsilon$. Similarly,
probabilistic estimates for the orbital period $P = 2\,\upi\,[a^3/(GM)]^{1/2}$ can be derived.

In contrast to the just 2 additional parameters $(d,q)$ as compared
to an ordinary microlensing light curves, our alternative channel
of detecting planets from the orbital motion of observed microlensing
source stars around the common barycentre involves
7 parameters $(\chi_\rmn{E},\psi,\Omega,i,\varphi,\varepsilon,t_\rmn{p})$.
For studies of the planet population, the values of the inclination $i$,
the phase angle $\varphi$, and the time of periastron $t_\rmn{p}$ are
of little use. Moreover, while for the annual parallax, the direction
angle of the source trajectory
$\psi$ with respect to the ecliptic plane, thereby being well-defined in
space, carries useful information, the direction angle $\psi$ with respect to the orbital plane of the planet does not.
In sharp contrast, a direct determination of 
the orbital period $P = (2\,\upi)/\Omega$ and the orbital eccentricity
$\varepsilon$ are valuable. 

Since spectral typing not only provides an estimate of the angular
source size $\theta_\star$, but also of the stellar mass $m_\star$,
the orbital semi-major axis $a = Gm [P^2/(4\,\upi^2)]^{1/3}$
is determined alongwith the orbital period $P$, given that
$m \approx m_\star$. The mass of the planet therefore follows
as $m_\rmn{p} = \chi_\rmn{E}\,[(D_\rmn{S}\,\theta_\rmn{E})/a]\,m$, where
the main indeterminacy results from the unknown $\theta_\rmn{E}$ with
in general only $t_\rmn{E} = \theta_\rmn{E}/\mu$ being extractable.
However, with finite-source effects being observed, $\theta_\rmn{E}$ 
is measured, and thereby the planet mass $m_\rmn{p}$ will result.
Given the fact that we do not require any knowledge about the lens
distance $D_\rmn{L}$, a measurement of the microlens parallax
$\pi_\rmn{E} = \pi_\rmn{LS}/\theta_\rmn{E}$ is not helpful in this case.
Otherwise, as for the 'standard' channel, one only finds a probability
density for $m_\rmn{p}$, but in contrast to that case, one obtains
measurements for the orbital semi-major axis $a$, the orbital
period $P$, and the orbital eccentricity $\varepsilon$.

While the parameters $(\chi_\rmn{E},\psi,\Omega,i,\varphi,\varepsilon,t_\rmn{p})$ are extractable in principle, the power for determining the properties of the planet and its orbit are limited by severe degeneracies that occur in several (not unlikely) cases.
For example, a proper measurement of the orbital period $P$ is only
possible for $P \la 3\,t_\rmn{E}$, whereas for wider orbits, the effect
on the microlensing light curve is mainly described by the 
acceleration of the source trajectory in the vicinity of the epoch
at which the ordinary, unperturbed light curve reaches its peak
\citep{smith02}. Moreover, for $t_\rmn{E} \la P \la 3\,t_\rmn{E}$,
the orbital period $P$ appears to be strongly degenerate with the
orbital inclination $i$ and cannot be properly disentangled.

Finally, with a large variety of effects causing small deviations, one
needs to take care to investigate all possible alternatives
such as lens or source binarity, and in particular the annual Earth-Sun
parallax. As already pointed out in Sect.~\ref{sec:reflex}, the latter
causes an identical signature, with the only difference that the
parameter space is restricted to the fixed values that define the Earth's
orbit.

\section{Summary and final conclusions}
\label{sec:conclusions}
We find that an alternative channel for the detection of extra-solar 
planets by microlensing is provided by the orbital motion of the source star around the common barycentre with an unseen planetary companion,
as opposed to the standard channel where a planet orbiting the lens
stars alters the bending angle and thereby the observed magnification.
We derived a formalism in exact analogy to the treatment of the annual
parallax that results from the revolution of the Earth around the Sun,
mainly following \citet{Do:rotate}, which however involves 7 additional
free parameters as compared to an ordinary microlensing light curve, which reduce to 5 for circular orbits, whereas there are
just 2 for the annual parallax due to the known properties of
the Earth's orbit. 

Constituting an exchange of the roles of source star and observer as
compared to parallax effects, which are most prominent for lens stars close to the observer, the strongest signals arise for lens stars close to the source star. In this limit, the signal strength practically depends
on the difference $D_\rmn{S}-D_\rmn{L}$ of the source and lens distances
only, rather than on the individual values, so that planetary signals
on observed Galactic bulge stars
will show up predominantly in events that involve lens stars in the
Galactic bulge rather than the Galactic disk. Moreover, the signal strength
persists for source stars in neighbouring galaxies,
such as M31, but achieving a sufficient photometric accuracy (on an
unresolved) target for being able to claim a detection is extremely 
challenging.

Other than for the standard microlensing channel, which is strongly biased in favour of K- and M-dwarfs due to their larger abundance,
more massive stars are the most prominent targets due to their greater
brightness, which results in favourable prospects for studying gas-giant 
planets, which are known to be extremely rare around low-mass stars.

Unless the orbital period can be identified from the observed light curve,
there are a variety of alternative explanations for the nature of the
event that are compatible with the acquired data. Rather featureless
deviations might also result from finite-source effects,
the revolution of the Earth around the Sun, as well as lens
or source binarity. Therefore, a proper 
characterization can only be expected if the orbital period $P$ does not
substantially exceed the event time-scale $t_\rmn{E}$, which works
against the fact that signals increase with $P$. Apart from this,
planets in closer orbits, in particular with $P \la 50~\mbox{d}$, could
be as well orbiting the lens star rather than the source star, given that
their effect on the bending of light would be expected to be negligible.

A rough estimate of the signal strength showed that the observability
of signals with current experimental setups is practically limited to
massive gas giants, and a Monte-Carlo simulation of a survey with
2~\mbox{h} sampling and 2 per cent photometric uncertainties revealed that,
for the Galactic bulge,
the detection of Jupiter-mass planets will be dominated by events
with small impact parameters $u_0 \la 0.03$, whereas there is a substantial chance
to detect planets of mass $m_\rmn{p} \sim 10~M_\rmn{jup}$ for the
larger sample of events
with $u_0 \la 0.1$. Microlensing observations with improved capabilities,
namely 15~min sampling with 0.3 per cent accuracy, would substantially 
increase the prospects, where the detection efficiency for Jupiter-mass
planets in events with $u_0 < 0.1$ being boosted from 0.9 per cent 
to 30 per cent, whereas the prospect for detecting planets that are 10
times more massive amongst all events with $u_0 < 1$ is pushed from
2.5 per cent to 50 per cent. With the much higher detection efficiencies
for moderate impact parameters, upper decades of $u_0$ (which carry
more events) would provide a larger fraction of the detections than 
lower decades.

If one is able to properly measure the orbital period $P$ from the
microlensing light curve, which requires $P \la t_\rmn{E}$, the
orbital semi-major axis $a$ can also be determined provided that
typing of the source star yields its mass $m_\star$. Along with the
orbital eccentricity $\varepsilon$, substantially more information can
be extracted as compared to the standard channel that lacks of 
vital constraints on the planetary orbit. The planet mass $m_\rmn{p}$
however still depends on the angular Einstein radius $\theta_\rmn{E}$,
which is not known in general, leaving the need to adopt a kinematic
model for the lens and source populations, as well as a mass function
for the lens stars, as discussed by \citet{Do:Esti2}. This can only
be overcome by additional measurements of either $\theta_\rmn{E}$
or the lens-source proper
motion $\mu$, e.g.\ from the assessment of finite-source effects
involving the time-scale $t_\star = \theta_\star/\mu$, where the
angular size of the source star $\theta_\star$ follows from typing
alongside the mass $m_\star$.

Current microlensing observing campaigns are not 
expected to provide many such detections, but with a 
detection probability of a few per cent for $m_\rmn{p} = 10~M_\rmn{jup}$,
and an estimated abundance also of a few per cent \citep{US:abundance}, a corresponding signal may already be present among the several thousand events that have been monitored so far (with 700-1000 new ones currently being discovered every year). However, finding it may require a careful extensive data-mining effort, given that deviations of this kind are easily missed or misidentified.


\section*{Acknowledgments}
We would like to thank Valerio Bozza, Scott Gaudi, Andy Gould, and
Cheongho Han for valuable comments and suggestions.

\bibliographystyle{mn2e}
\bibliography{srcplanet}

\appendix
\section{Models of the Galactic bulge population}
\label{app:bulge}
Given that the number of source stars in the observed field is proportional to $\rho_\rmn{S}(D_\rmn{S})\,D_\rmn{S}^2$, while the event rate for
a given source star is proportional to $\rho_\rmn{L}(D_\rmn{L})\,D_\rmn{L}\,\theta_\rmn{E}$, where $\rho_\rmn{S}$ and $\rho_\rmn{L}$ denote 
the volume mass densities of the source and lens stars, respectively,
a joint probability density for the source and lens distances is
proportional to the differential total event rate
\begin{eqnarray}
\frac{\rmn{d}^2\Gamma}{\rmn{d}D_\rmn{L}\,\rmn{d}D_\rmn{S}}
 & \propto & \rho_\rmn{L}(D_\rmn{L})\,\rho_\rmn{S}(D_\rmn{S})\,\times \nonumber\\
& & \hspace*{-5em} \times\,D_\rmn{S}^2\,
\sqrt{D_\rmn{L}\,\left(1-\frac{D_\rmn{L}}{D_\rmn{S}}\right)}\;\Theta(D_\rmn{S}-D_\rmn{L})\,\Theta(D_\rmn{L})
\label{eq:jointDLDS}
\end{eqnarray}
with $\Theta(x)$ denoting the step function. This however does not
account for the fact that more distant source stars appear fainter on average and that their light is more likely to be affected by extinction.
Integration of Eq.~(\ref{eq:jointDLDS}) over the source distance $D_\rmn{S}$ leads to the probability density of the lens distance $D_\rmn{L} > 0$ being
proportional to
\begin{eqnarray}
\frac{\rmn{d}\Gamma}{\rmn{d}D_\rmn{L}}
 & \propto &  \nonumber\\
& & \hspace*{-5em} \propto\,\rho_\rmn{L}(D_\rmn{L})\,\sqrt{D_\rmn{L}}\,\int\limits_{D_\rmn{L}}^{\infty}\rho_\rmn{S}(D_\rmn{S})\,D_\rmn{S}^2\,
\sqrt{1-\frac{D_\rmn{L}}{D_\rmn{S}}}\;\rmn{d}D_\rmn{S}\,.
\label{eq:probDL}
\end{eqnarray}
Alternatively, the joint probability density as given by Eq.~(\ref{eq:jointDLDS}) can be transformed to refer to the source-lens distance $D_\rmn{LS} \equiv D_\rmn{S}-D_\rmn{L}$ rather than
the lens distance $D_\rmn{L}$, yielding
\begin{eqnarray}
\frac{\rmn{d}^2\Gamma}{\rmn{d}D_\rmn{LS}\,\rmn{d}D_\rmn{S}}
 & \propto & \rho_\rmn{L}(D_\rmn{S}-D_\rmn{LS})\,\rho_\rmn{S}(D_\rmn{S})\,\times \nonumber \\
& & \hspace*{-5em} \times\,D_\rmn{S}^2\,
\sqrt{D_\rmn{LS}\,\left(1-\frac{D_\rmn{LS}}{D_\rmn{S}}\right)}\;\Theta(D_\rmn{LS})\,\Theta(D_\rmn{S}-D_\rmn{LS})\,.
\label{eq:jointDLSDS}
\end{eqnarray}
Therefore, one finds the respective probability density of $D_\rmn{LS}$
as
\begin{eqnarray}
\frac{\rmn{d}\Gamma}{\rmn{d}D_\rmn{LS}}
 & \propto & \sqrt{D_\rmn{LS}}\,\int\limits_{D_\rmn{LS}}^{\infty} \rho_\rmn{L}(D_\rmn{S}-D_\rmn{LS})\,\rho_\rmn{S}(D_\rmn{S})\,\times \nonumber \\
& & \hspace*{7em} \times\,D_\rmn{S}^2\,
\sqrt{1-\frac{D_\rmn{LS}}{D_\rmn{S}}}\;\rmn{d}D_\rmn{S}\,.
\label{eq:probDLS}
\end{eqnarray}

Amongst the various laws suggested for the mass density of the Galactic
bulge, we adopt a triaxial barred shape of the form
\begin{equation}
\rho_\rmn{bulge} = \frac{M_\rmn{bulge}}{6.57\pi s_x s_y s_z }\;\exp\left(-r^2/2\right)\,,
\label{bulge_dist}
\end{equation}
where 
\begin{equation}
r^4=\left[(\frac{\hat x}{s_x})^2 +(\frac{\hat y}{s_y})^2\right]^2 +
(\frac{\hat z}{s_z})^4\,,
\end{equation} 
which is a favourite for matching COBE data \citep{Dwe95} with
$s_x=1.49~\mbox{kpc}$, $s_y=0.58~\mbox{kpc}$ and $s_z=0.40~\mbox{kpc}$,
and $M_\rmn{bulge} = 1.7\times 10^{10}~M_{\odot}$. 
While $\hat z$ measures the distance towards Galactic North,
the axes referring to the $\hat x$ and $\hat y$ coordinates are
tilted by an angle $\theta \sim 20\degr$ with respect to the 
direction towards the Galactic centre and the direction of the local
circular motion, respectively.

We neglect the variation with the sky coordinates of the source star, and
adopt a typical average Galactic longitude $l = 0\degr$ and
latitude $b= -3\degr$,
so that
\begin{eqnarray}
\hat x &=& (D_L\cos{b}-R_\rmn{GC})\cos{\theta}\, \nonumber \\
\hat y &=& -(D_L\cos{b}-R_\rmn{GC})\sin{\theta}\, \nonumber \\
\hat z &=&D_L\sin{b}\,,
\end{eqnarray}
where $R_\rmn{GC} \sim 7.62~\mbox{kpc}$ denotes the distance to the
Galactic centre.

\begin{figure}
\begin{center}
\includegraphics[width=84mm]{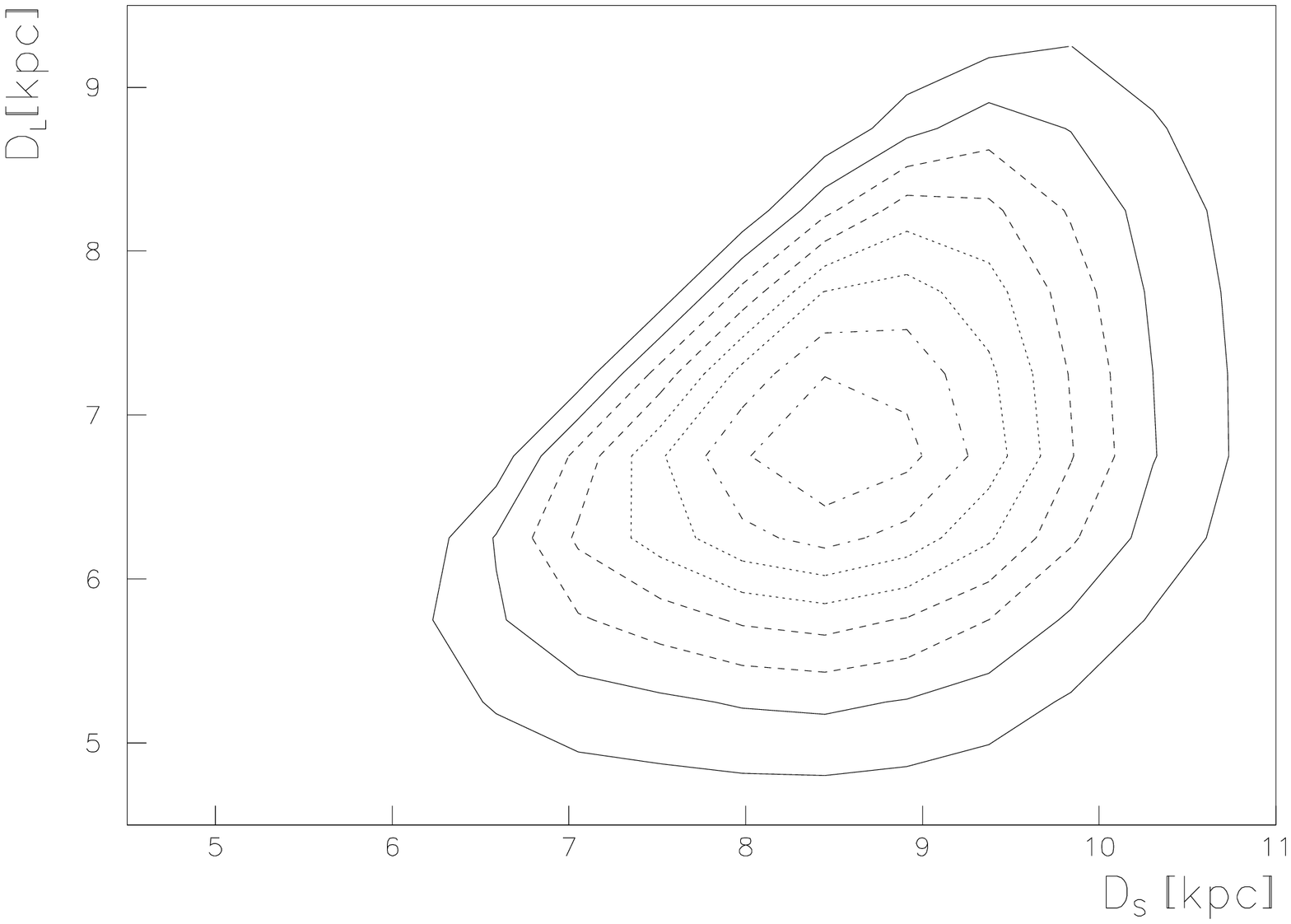}
\caption{Joint probability density $\Gamma^{-1} [\rmn{d}^2\Gamma/(\rmn{d}D_\rmn{L}\,\rmn{d}D_\rmn{S})]\,(D_\rmn{L}, D_\rmn{S})$, indicated by equidistant contours between zero and the maximum, of the
lens distance $D_\rmn{L}$ and the source distance $D_\rmn{S}$ for
self-lensing withing the Galactic bulge in the direction of $l=0\degr$ and $b=-3\degr$.}
\label{fig:jointprob}
\end{center}
\end{figure}

\begin{figure}
\begin{center}
\includegraphics[width=84mm]{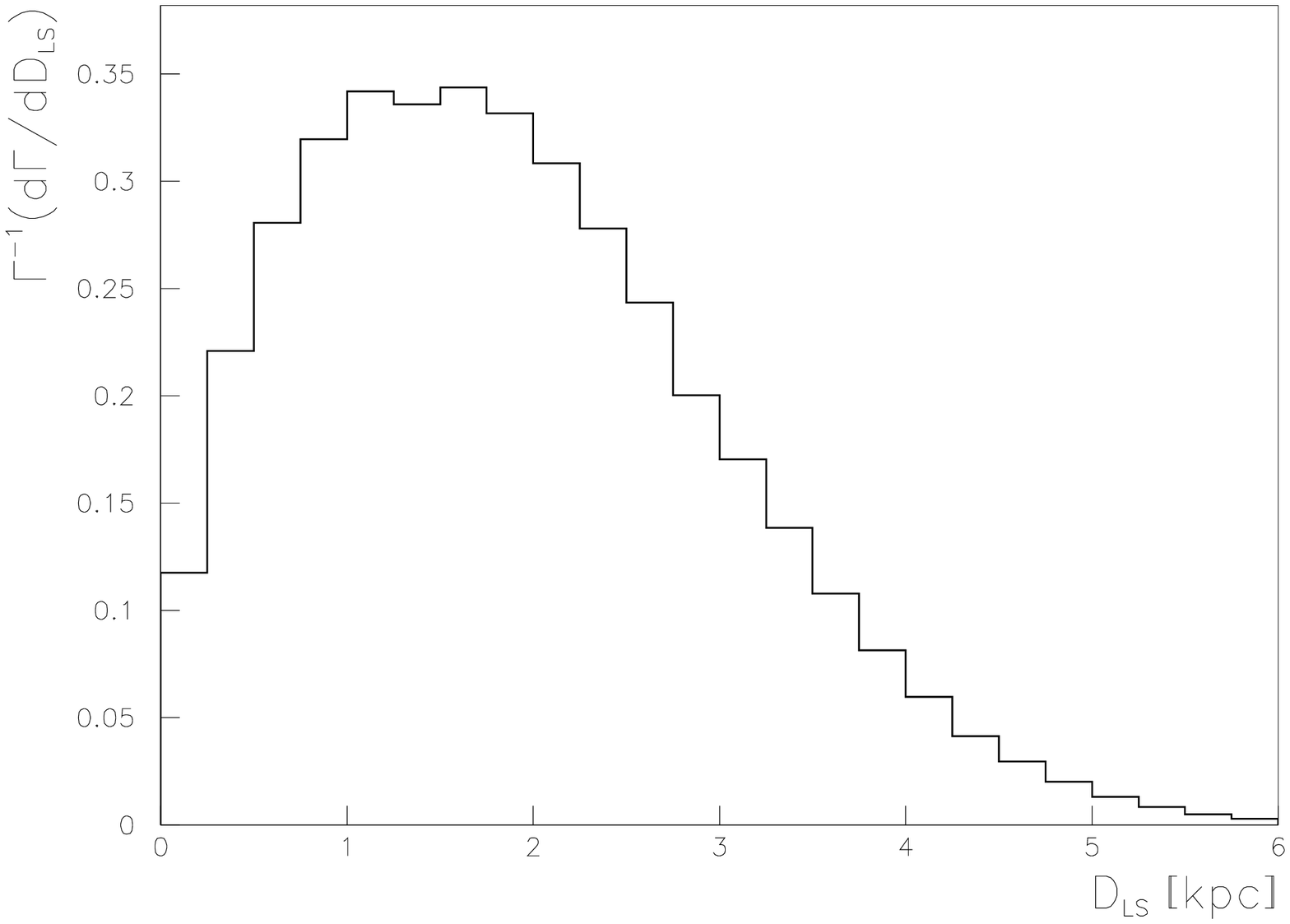}
\caption{Probability density of the lens-source distance $\Gamma^{-1}\,(\rmn{d}\Gamma/\rmn{d}D_\rmn{LS})\,(D_\rmn{LS})$ for bulge-bulge lensing, where the source stars have been located
towards $l=0\degr$ and $b=-3\degr$ and the bulge mass density is given by
Eq.~(\ref{bulge_dist}).} 
\label{fig:dlsrate}
\end{center}
\end{figure}

Given that the bulge self-lensing is dominant, $\rho_\rmn{bulge}$ 
describes both the lens and the source stars. The joint probability 
density of the lens distance $D_\rmn{L}$ and source distance $D_\rmn{S}$,
according to Eq.~(\ref{eq:jointDLDS}), as well as the probability 
density of the lens-source distance $D_\rmn{LS} 
\equiv D_\rmn{S}-D_\rmn{L}$, according to Eq.~(\ref{eq:probDLS}), respectively, that result from this choice,
are displayed in Figs.~\ref{fig:jointprob} and~\ref{fig:dlsrate}.

For the velocities of lens and source stars,
we adopt a two-dimensional Maxwell-Boltzmann distribution in the 
absolute transversal velocity $v$, where the probability to find $v$ in the
interval $[v,v+\rmn{d}v]$ is given by
\begin{equation}
\Phi_v(v)\,\rmn{d}v = \frac{v}{2\,\sigma^2}\,\exp\left(
-\frac{v^2}{4\,\sigma^2}\right)\,\rmn{d}v\,.
\end{equation}
While we adopted $\sigma = 100~\mbox{km}\,\mbox{s}^{-1}$, we
neglected its effective variation with $D_\rmn{L}$ and $D_\rmn{S}$, as well as the motion of the Sun,
which are fair approximations for $D_\rmn{L}/D_\rmn{S} \sim 1$,
marking the region dominating the event count.

\begin{figure}
\begin{center}
\includegraphics[width=84mm]{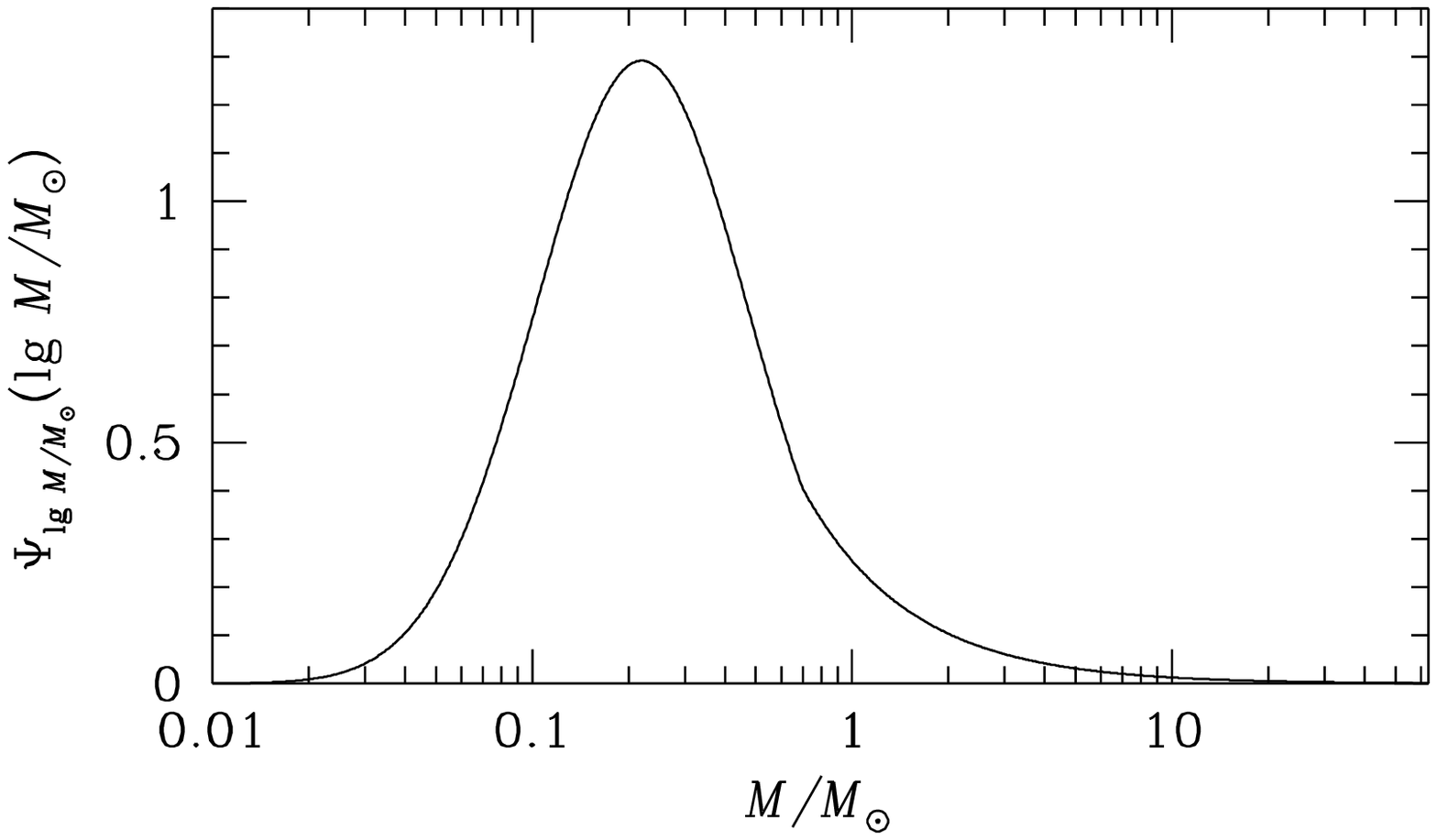}
\caption{The adopted normalized Galactic bulge mass function
$\Psi_{\lg (M/M_{\sun})}[\lg (M/M_{\sun})]$, which corresponds to
the probability density in $\lg (M/M_{\sun})$ for the mass of a
given star.}
\label{fig:bulgemass}
\end{center}
\end{figure}

For the mass of Galactic bulge stars, a probability density in $\lg (M/M_{\sun})$ is given by the normalized mass function \citep*[e.g.][]{Cha03}
\begin{eqnarray}
\Psi_{\lg (M/M_{\sun})}[\lg (M/M_{\sun})] &  = &  \nonumber \\
& & \hspace*{-14em} =\, \left\{ \begin{array}{l} 1.292\,\exp\left\{-0.5\left[
\frac{\lg(M/M_{\sun})+0.658}{0.33}\right]^2\right\} \\ 
\hspace*{5em} \mbox{for}\; -2.0 \leq \lg (M/M_{\sun}) \leq -0.155
\\ 0.2546\,(M/M_{\sun})^{-1.3}\\
\hspace*{5em} \mbox{for}\; -0.155 < \lg (M/M_{\sun}) \leq 1.8 \\
 0 \quad \mbox{for}\; \lg (M/M_{\sun}) < -2.0 \; \mbox{or}\;
              \lg (M/M_{\sun}) > 1.8\end{array} \right. \,,
\end{eqnarray}
which is shown in Figure~\ref{fig:bulgemass}.

\end{document}